\documentclass[fleqn,usenatbib]{mnras}

\usepackage[T1]{fontenc}
\usepackage{ae,aecompl}

\usepackage{graphicx}	% Including figure files
\usepackage{amsmath}	% Advanced maths commands
\usepackage{amssymb}	% Extra maths symbols

\usepackage{longtable}
\usepackage{lscape}
\usepackage{siunitx} 
\usepackage{subfig}
\usepackage{threeparttable}

\graphicspath{{Images/}}

\usepackage{eso-pic}% http://ctan.org/pkg/eso-pic

\newcommand{\mstellar}{\ensuremath{M_\mathrm{stellar}}}

\usepackage{mwe}

\usepackage{newtxtext,newtxmath}

\defcitealias{2020MNRAS.495.4040W}{W20}
\defcitealias{2020MNRAS.494.4426S}{S20}

\title[The effect of environment on SNe Ia]{The Effect of Environment on Type Ia Supernovae in the Dark Energy Survey Three-Year Cosmological Sample}

\author[L. Kelsey et al.]{
\parbox{\textwidth}{
\Large
L.~Kelsey,$^{1,2}$\thanks{E-mail: l.kelsey@soton.ac.uk}
M.~Sullivan,$^{1}$
M.~Smith,$^{1,3}$
P.~Wiseman,$^{1}$
D.~Brout,$^{4,5,6}$
T.~M.~Davis,$^{7}$
C.~Frohmaier,$^{8}$
L.~Galbany,$^{9}$
M.~Grayling,$^{1}$
C.~P.~Guti\'errez,$^{1}$
S.~R.~Hinton,$^{7}$
R.~Kessler,$^{10,11}$
C.~Lidman,$^{12,13}$
A.~M\"oller,$^{14}$
M.~Sako,$^{4}$
D.~Scolnic,$^{15}$
S.~A.~Uddin,$^{16}$
M.~Vincenzi,$^{8}$
T.~M.~C.~Abbott,$^{17}$
M.~Aguena,$^{18,19}$
S.~Allam,$^{20}$
J.~Annis,$^{20}$
S.~Avila,$^{21}$
D.~Bacon,$^{8}$
E.~Bertin,$^{22,23}$
D.~Brooks,$^{24}$
D.~L.~Burke,$^{25,26}$
A.~Carnero~Rosell,$^{27,28}$
M.~Carrasco~Kind,$^{29,30}$
J.~Carretero,$^{31}$
F.~J.~Castander,$^{32,33}$
M.~Costanzi,$^{34,35}$
L.~N.~da Costa,$^{19,36}$
S.~Desai,$^{37}$
H.~T.~Diehl,$^{20}$
P.~Doel,$^{24}$
S.~Everett,$^{38}$
I.~Ferrero,$^{39}$
A.~Fert\'e,$^{40}$
B.~Flaugher,$^{20}$
P.~Fosalba,$^{32,33}$
J.~Garc\'ia-Bellido,$^{21}$
D.~W.~Gerdes,$^{41,42}$
D.~Gruen,$^{43,25,26}$
R.~A.~Gruendl,$^{29,30}$
J.~Gschwend,$^{19,36}$
G.~Gutierrez,$^{20}$
D.~L.~Hollowood,$^{38}$
K.~Honscheid,$^{44,45}$
D.~J.~James,$^{6}$
A.~G.~Kim,$^{46}$
K.~Kuehn,$^{47,48}$
N.~Kuropatkin,$^{20}$
O.~Lahav,$^{24}$
M.~Lima,$^{18,19}$
J.~L.~Marshall,$^{49}$
P.~Martini,$^{44,50,51}$
F.~Menanteau,$^{29,30}$
R.~Miquel,$^{52,31}$
R.~Morgan,$^{53}$
R.~L.~C.~Ogando,$^{19,36}$
A.~Palmese,$^{20,11}$
F.~Paz-Chinch\'{o}n,$^{54,30}$
A.~A.~Plazas,$^{55}$
A.~K.~Romer,$^{56}$
C.~S{\'a}nchez,$^{4}$
E.~Sanchez,$^{57}$
S.~Serrano,$^{32,33}$
I.~Sevilla-Noarbe,$^{57}$
E.~Suchyta,$^{58}$
G.~Tarle,$^{42}$
D.~Thomas,$^{8}$
C.~To,$^{43,25,26}$
T.~N.~Varga,$^{59,60}$
A.~R.~Walker,$^{17}$
and R.D.~Wilkinson$^{56}$
\begin{center} (DES Collaboration) \end{center}
}
\vspace{0.4cm}
\\
\parbox{\textwidth}{Affiliations are listed at the end of the paper}
}

\date{Accepted XXX. Received YYY; in original form ZZZ}

\pubyear{2020}

\begin{document}
\label{firstpage}
\pagerange{\pageref{firstpage}--\pageref{lastpage}}
\maketitle

\begin{abstract}
Analyses of type Ia supernovae (SNe Ia) have found puzzling correlations between their standardised luminosities and host galaxy properties: SNe Ia in high-mass, passive hosts appear brighter than those in lower-mass, star-forming hosts. We examine the host galaxies of SNe Ia in the Dark Energy Survey three-year spectroscopically-confirmed cosmological sample, obtaining photometry in a series of \lq local\rq\ apertures centred on the SN, and for the global host galaxy. We study the differences in these host galaxy properties, such as stellar mass and rest-frame $U-R$ colours, and their correlations with SN Ia parameters including Hubble residuals. We find all Hubble residual steps to be $>3\sigma$ in significance, both for splitting at the traditional environmental property sample median and for the step of maximum significance. For stellar mass, we find a maximal local step of $0.098\pm0.018$\,mag; $\sim 0.03$\,mag greater than the largest global stellar mass step in our sample ($0.070 \pm 0.017$\,mag). When splitting at the sample median, differences between local and global $U-R$ steps are small, both $\sim 0.08$\,mag, but are more significant than the global stellar mass step ($0.057\pm0.017$\,mag). We split the data into sub-samples based on SN Ia light curve parameters: stretch ($x_1$) and colour ($c$), finding that redder objects ($c > 0$) have larger Hubble residual steps, for both stellar mass and $U-R$, for both local and global measurements, of $\sim0.14$\,mag. Additionally, the bluer (star-forming) local environments host a more homogeneous SN Ia sample, with local $U-R$ r.m.s. scatter as low as $0.084 \pm 0.017$\,mag for blue ($c < 0$) SNe Ia in locally blue $U-R$ environments. 
\end{abstract}


%%%%%%%%%%%%%%%%%%%%%%%%%%%%%%%%%%%%%%%%%%%%%%%%%%

%%%%%%%%%%%%%%%%% BODY OF PAPER %%%%%%%%%%%%%%%%%%

\section{Introduction} \label{intro}

Type Ia supernovae (SNe Ia) are important cosmological probes due to their role as distance indicators, and most famously have been used to reveal the accelerating expansion of the universe \citep{1998AJ....116.1009R,  1999ApJ...517..565P}. Their low intrinsic peak absolute magnitude dispersion of $\sim$0.35\,mag can be standardised using \lq{brighter-slower}\rq\ \citep{1993ApJ...413L.105P} and \lq{brighter-bluer}\rq\ \citep{1996ApJ...473...88R, 1998A&A...331..815T} relations to achieve a $\sim$0.14\,mag dispersion \citep{2018ApJ...859..101S}.
While some of this remaining scatter can be attributed to observational uncertainties, there remains an \lq intrinsic dispersion\rq\ of $\simeq$0.08-0.10\,mag \citep{2019ApJ...874..150B}. This indicates either the
limit to which SNe Ia are standardisable, or that there are further brightness correlations that cannot be uncovered with the size and quality of current samples. This latter possibility could arise from astrophysical uncertainties in the SN Ia progenitor mechanisms, explosion physics, and/or environment \citep{2014ARA&A..52..107M,Maguire2017,2018PhR...736....1L}.

The desire for an improved standardisation for SNe Ia has motivated more than 20 years of work searching for correlations between the properties of SNe Ia and the closest proxy we have for their progenitor stellar populations: their host galaxies. There is strong evidence that the colour- and stretch-corrected brightness correlate with the stellar mass of the SN Ia host galaxy: SNe Ia in high-mass hosts standardise to brighter luminosities than those in lower-mass hosts \citep[e.g.][]{2010MNRAS.406..782S, 2010ApJ...715..743K, 2010ApJ...722..566L}. The stellar mass of galaxies correlates with the stellar ages, gas-phase and stellar metallicities, and dust content of its stellar populations \citep{2004ApJ...613..898T, 2005MNRAS.362...41G,2010MNRAS.409..421G,2011MNRAS.414.1592B,2013ApJ...763...92Z}, suggesting that the trends between corrected SN Ia brightness and host stellar mass could be due to differences in intrinsic SN progenitor properties \citep[e.g., age or metallicity;][]{2003ApJ...590L..83T,2004A&A...420L...1R,2009Natur.460..869K,2010ApJ...711L..66B} or dust \citep[e.g.,][]{2020arXiv200410206B}. The physical nature of the dominant underlying effect remains controversial. 

Nonetheless, the empirical dependence of the corrected Hubble residuals on the SN Ia host galaxy stellar mass has now been studied extensively \citep[e.g.,][]{2011ApJ...740...92G,2013MNRAS.435.1680J,2013ApJ...770..108C, 2017ApJ...848...56U, 2020MNRAS.494.4426S}, including in the near infrared \citep{2020arXiv200613803P,2020ApJ...901..143U}, with modern samples having evidence for a step in calibrated SN Ia magnitude of $\sim0.06$\,mag at around $\log(\mstellar/M_{\sun})\simeq 10$, where \mstellar\ is the stellar mass of the SN host galaxy. Studies have also extended to other galaxy properties such as specific star-formation rate (sSFR; the star-formation rate per unit stellar mass) and metallicity, with similar steps in SN corrected luminosity being observed \citep[e.g.,][]{2013A&A...560A..66R}.  

Focusing on the \lq local\rq\ host galaxy properties at the SN Ia position, rather than the global properties of the host galaxy, perhaps provides a more immediate census of the stellar populations from which the progenitor was drawn \citep{2013A&A...560A..66R, 2015ApJ...802...20R, 2015ApJ...812...31J, 2018ApJ...867..108J, 2018A&A...615A..68R, 2018ApJ...855..107G, 2019ApJ...874...32R, 2018ApJ...854...24K, 2019JKAS...52..181K}. Global galaxy properties, such as the star formation rate (SFR), are weighted by surface brightness, meaning that global measurements are most representative of the properties of the brightest galactic regions, and thus may not be accurate measurements of the true environment of the progenitor and resulting SN \citep{2013A&A...560A..66R}. On the other hand, any correlation with local host properties are diluted if the birth place of the progenitor differs from the region the SN explodes, an effect that becomes larger with longer delay times. By using data from the Nearby Supernova Factory \citep{2002SPIE.4836...61A} to measure nebular H$\alpha$ emission from \ion{H}{ii} regions (as a tracer of SFR), \cite{2013A&A...560A..66R} found correlations between the local SFR within a 1\,kpc radius around each SN and the SN Ia corrected magnitude, in which SNe Ia in locally star-forming environments are fainter than those in locally passive environments by $\sim0.094$\,mag. This relationship was later confirmed using the \lq Constitution\rq\ SN sample \citep{2009ApJ...700.1097H} combined with \textit{GALEX} host galaxy data \citep{2015ApJ...802...20R}. 

\cite{2018A&A...615A..68R} analysed to higher redshift using the local rest-frame $U-V$ colour of the host galaxy at the SN Ia position in place of H$\alpha$, using a compilation of Supernova Legacy Survey \citep[SNLS;][]{2006A&A...447...31A} 5-year data, the Sloan Digital Sky Survey \citep[SDSS;][]{2018PASP..130f4002S} SN survey, and various low-redshift surveys. The step in corrected magnitude from blue to red environments was $0.091\pm0.013$\,mag, comparable to the global galaxy mass step found by \cite{2013ApJ...770..108C}. This step persists when a correction for the mass step is performed first, although decreases to $0.057\pm0.012$\,mag. Using a larger low-redshift SN Ia sample \citep[including SNe Ia from the \lq Foundation\rq\ sample;][]{2018MNRAS.475..193F}, \cite{2018ApJ...867..108J} found similar-sized steps to \citet{2018A&A...615A..68R} using local stellar mass and local $u-g$ colour, although at lower significance. A further nuance was that low-redshift SNe Ia discovered in targeted galaxy surveys showed no local stellar mass or colour steps, while SNe Ia located in the \lq rolling\rq\ Foundation survey similar to SNLS or SDSS showed a significant local step.

\cite{2018arXiv180603849R} developed these ideas further by statistically classifying a sample of SNe Ia from the Nearby Supernova Factory into younger or older environments based on the local specific star formation rate (LsSFR) measured within a distance of 1\,kpc from each SN. They found that SNe in younger environments are fainter at 5.7$\sigma$ significance than those in older environments after light-curve correction. As the average age of stellar populations evolve with redshift, this could create a bias in cosmological analysis. In a recent study using global properties to infer local environmental properties, \cite{2019JKAS...52..181K} suggest this environmental dependence of SNe Ia may lead to an evolution in their mean luminosity with redshift.

In this paper we use data from the Dark Energy Survey \citep[DES;][]{2016MNRAS.460.1270D} first three-year cosmological sample \citep[DES3YR;][]{2019ApJ...874..150B} to measure the correlation between SN~Ia luminosity and local environment. We combine this sample with photometry based on deep stacks of optical data free from SN light \citep[hereafter W20]{2020MNRAS.495.4040W}, we measure the local stellar mass and colour for each SN~Ia in a range of physical aperture sizes. This dataset spans a wide redshift range ($0.02<z<0.8$) with SN candidates identified and spectroscopically targeted using algorithms principally agnostic to local environment \citep{2015AJ....150..172K,2020AJ....160..267S}. When analysed using the \lq BEAMS with Bias Corrections\rq\ \citep[BBC;][]{2017ApJ...836...56K} framework, the DES-SN sample finds evidence of a correlation between global host stellar mass and Hubble residuals consistent with literature samples, but dependent on the bias correction considered \citep[hereafter S20]{2020MNRAS.494.4426S}. This paper builds on the work of \citetalias{2020MNRAS.494.4426S}, not only looking at the host galaxy stellar mass along the line of sight, but additionally primarily focusing on the rest-frame $U-R$ colour, and studying the effects of host galaxy environmental properties at the local scale.  

The paper has the following structure. In Section~\ref{data} we describe the DES-SN programme and the sample that was used for this study, before presenting the method used to obtain both global and local host galaxy aperture photometry used to study the environmental dependence of SN luminosity, the results of which are discussed in Section~\ref{results}. In Section~\ref{sys} we present the additional tests used to test the robustness of our analysis, and conclude in Section~\ref{discussion} by putting this work into context with previous studies. We assume a spatially-flat $\Lambda$CDM model, with a matter density $\Omega_m = 0.3$ and Hubble constant $H_0 = 70 \textrm{km s}^{-1} \textrm{Mpc}^{-1}$, and use AB magnitudes \citep{1983ApJ...266..713O} throughout. 

\section{Data and methods}
\label{data}

We begin with a description of the SN Ia sample that we use in this study, and the associated data on their host and local environments. Next, we describe our measurements and the host galaxy parameters that we calculate from these data.

\subsection{The DES-SN Ia sample}
\label{SNsample}

DES was a six-year imaging survey covering $\sim5100$ square degrees of the southern hemisphere using the 4-m Blanco telescope at the Cerro Tololo Inter-American Observatory (CTIO), equipped with the 520 megapixel wide-field Dark Energy Camera \citep[DECam;][for an overview]{2015AJ....150..150F} with a 0.263 arcsecond per pixel resolution. The survey included a five-year transient survey (\lq DES-SN\rq), optimised for the detection and measurement of SNe Ia for cosmology.

DES-SN was designed to obtain several thousand SN Ia light curves over $0.2 < z < 1.2$ \citep{2012ApJ...753..152B, 2020AJ....160..267S}, with eight \lq{shallow}\rq\ fields (E1, E2, S1, S2, C1, C2, X1, X2; with an average single-epoch depth of $23.5$\,mag\footnote{Where depth here refers to the magnitude at which 50\% of artificially injected point sources are recovered, see \citet{2015AJ....150..172K}.}) and two \lq{deep}\rq\ fields (C3 and X3; with an average single-epoch depth of $24.5$\,mag), each a single DECam pointing, finding SNe Ia at both intermediate and high redshift. DES-SN observed each field in the \textit{griz} filters with a mean cadence of $\simeq7$ days. The data were processed by the DES Data Management team \citep[DESDM;][]{2018PASP..130g4501M} for routine image detrending, and then processed by the DES-SN Difference Imaging Pipeline \citep[\textsc{DiffImg};] []{2015AJ....150..172K} to identify transient events.

We use the 206 spectroscopically-confirmed SNe Ia in the DES3YR sample that satisfy selection requirements (cuts) and were used in the cosmological analysis, covering a redshift range of $0.017 < z < 0.85$. Data from the full data release \footnote{\url{https://www.darkenergysurvey.org/des-year-3-supernova-cosmology-results/}} of the DES3YR SN Ia light curves and spectra, can be found as follows: cosmology sample and systematics in \citet{2019ApJ...874..106B}, photometry in \citet{2019ApJ...874..150B}, spectroscopy in  \citet{2020AJ....160..267S}.

\subsubsection{SN derived parameters} \label{params}

Each DES SN Ia light curve is fit with the SALT2 model \citep{2007A&A...466...11G, 2010A&A...523A...7G}, trained with the Joint Lightcurve Analysis compilation \citep[JLA;][]{2014A&A...568A..22B}, implemented in the \textsc{snana} software package \citep{2009PASP..121.1028K}. This fit returns a \lq stretch\rq\ ($x_1$) and \lq colour\rq\ ($c$) measurement for each SN Ia event, as well as the observed apparent magnitude ($m_B$). We calculate the \lq Hubble residual\rq\ for each SN, defined as the difference between the measured distance modulus ($\mu_\mathrm{obs}$) to each event, and the distance modulus calculated from the best-fit cosmology to the SN sample ($\mu_{\mathrm{cosmo}}$), i.e.,
\begin{equation}
    \Delta\mu = \mu_\mathrm{obs} - \mu_\mathrm{cosmo},
\label{equation:resi}
\end{equation}
where $\mu_\mathrm{obs}$ is defined as
\begin{equation}
    \mu_\mathrm{obs} = m_B - M_0 + \alpha x_1 - \beta c + \mu_\mathrm{bias},
\label{equation:hubble}
\end{equation}
and $\alpha$, $\beta$ and $M_0$ are nuisance parameters describing the SN population determined in the cosmological fit. In this analysis we use cosmological nuisance parameters from the DES3YR SN Ia analysis ($\alpha = 0.156 \pm 0.012$, $\beta = 3.201 \pm 0.131$), but study the effect of using a 5D correction in Section~\ref{vary_cosmo} and of refitting $\alpha$ and $\beta$ in Section~\ref{split_env_refit}.

The $\mu_\mathrm{bias}$ term is a bias correction, determined from the BBC method \citep{2019MNRAS.485.1171K} which makes use of simulations, made to each SN Ia to account for various survey selection effects. This correction is defined either as a \lq 1D correction\rq\ as a function of redshift, or as a \lq 5D correction\rq\ as a function of \{$z, x_1, c, \alpha, \beta$\} \citep{2017ApJ...836...56K}. \citetalias{2020MNRAS.494.4426S} showed that in the DES 3YR sample, a 1D bias correction gives a statistically-significant mass step of $0.066\pm0.020$\,mag, consistent with previous results. However, with the 5D bias correction, only a small mass step was found ($0.040\pm0.019$\,mag -- a difference of $0.026\pm0.009$\,mag). \citetalias{2020MNRAS.494.4426S} show this difference is likely due to an underlying correlation between host-galaxy stellar mass and SN Ia stretch that is not accounted for in current bias simulations. In this paper we therefore employ the 1D bias correction method (however, see Section~\ref{vary_cosmo} for a discussion on the 5D correction).

These mass steps lead to a further \lq host galaxy\rq\ correction in typical cosmological analyses, $\gamma G_\textrm{host}$,  where $G_\textrm{host}=\pm1/2$ and the sign depends on the value of a SN Ia host galaxy property, and $\gamma$ is analogous to $\alpha$ and $\beta$. This step function changes sign at some value of the SN host global property, which we label as the \lq division point\rq. For example, when using stellar mass,
\begin{equation}
    G_\textrm{host} = 
    \begin{cases}
    +1/2, & \text{if }\log\left(\mstellar/M_{\sun}\right) > M_{\textrm{step}} \\
    -1/2, & \text{if}\  M_{\textrm{step}} < \log\left(\mstellar/M_{\sun}\right)
    \end{cases}
\label{equation:ghost}
\end{equation}
where $M_{\textrm{step}}$ is the division point. In this analysis we do not fit for $\gamma$, instead we calculate $\mu_{\textrm{cosmo}}$ without the mass step to test physics and the potential cause of the SN Ia residual dispersion by studying \mstellar\ and the rest-frame $U-R$ colour in order to infer $\gamma$.

In most previous studies, $M_\textrm{step}$ was chosen to be at the median or mean stellar mass of the SN Ia sample, or arbitrarily chosen at some location \citep[e.g., $10^{10}\,\mathrm{M}_{\sun}$;][]{2010MNRAS.406..782S}. There is little physical motivation for this choice, although we note that $10^{10}\,\mathrm{M}_{\sun}$ lies just below the knee in the galaxy-mass/halo-mass relation \citep[$\sim3 x 10^{10}\,\mathrm{M}_{\sun}$\ at low redshift,][]{2003MNRAS.341...54K}, the point at which galaxies transform from \lq{star-formation-dominated SN-regulated}\rq, to \lq{accretion-dominated AGN-regulated}\rq\ growth \citep{2011IAUS..277..273S, 2013ApJ...772..112S,2017MNRAS.469.4249T, 2020MNRAS.491..634G}. This galaxy--halo connection is known to have effects on galaxy properties \citep[for a review, see][]{2018ARA&A..56..435W}. 

\subsection{Host galaxy measurements}

In this section we discuss the framework and methods used to obtain the photometric data for use in our analysis.

\subsubsection{Image stacking}
\label{stacks}

Analyses based on global host properties require deep imaging to reduce statistical uncertainties on the galaxy photometry. However, here we are interested in the local galaxy properties at the SN location, we require stacks that i) are not contaminated with light from the SN; and ii) have been optimised for seeing, as measured by the full-width half-maximum (FWHM) of the point spread function (PSF) of stars measured in the images. The signal-to-noise of galaxy photometry is shown to be significantly improved by coadding images, see e.g. \citetalias{2020MNRAS.495.4040W} Figure 8.  Our new image stacks follow the techniques of \citetalias{2020MNRAS.495.4040W}, and we build different stacks for each survey season of SN discovery, omitting data from the season of discovery in each case. Whilst we use the supernova sample from the first three years of the survey in this analysis (see e.g. \citealt{2018ApJS..239...18A,2019ApJ...874..106B,2019ApJ...874..150B}), we have five years of imaging available, which we utilise as per \citetalias{2020MNRAS.495.4040W} by considering imaging data taken across the five years of the DES survey.

\begin{figure}
\begin{center}
\includegraphics[width=\linewidth]{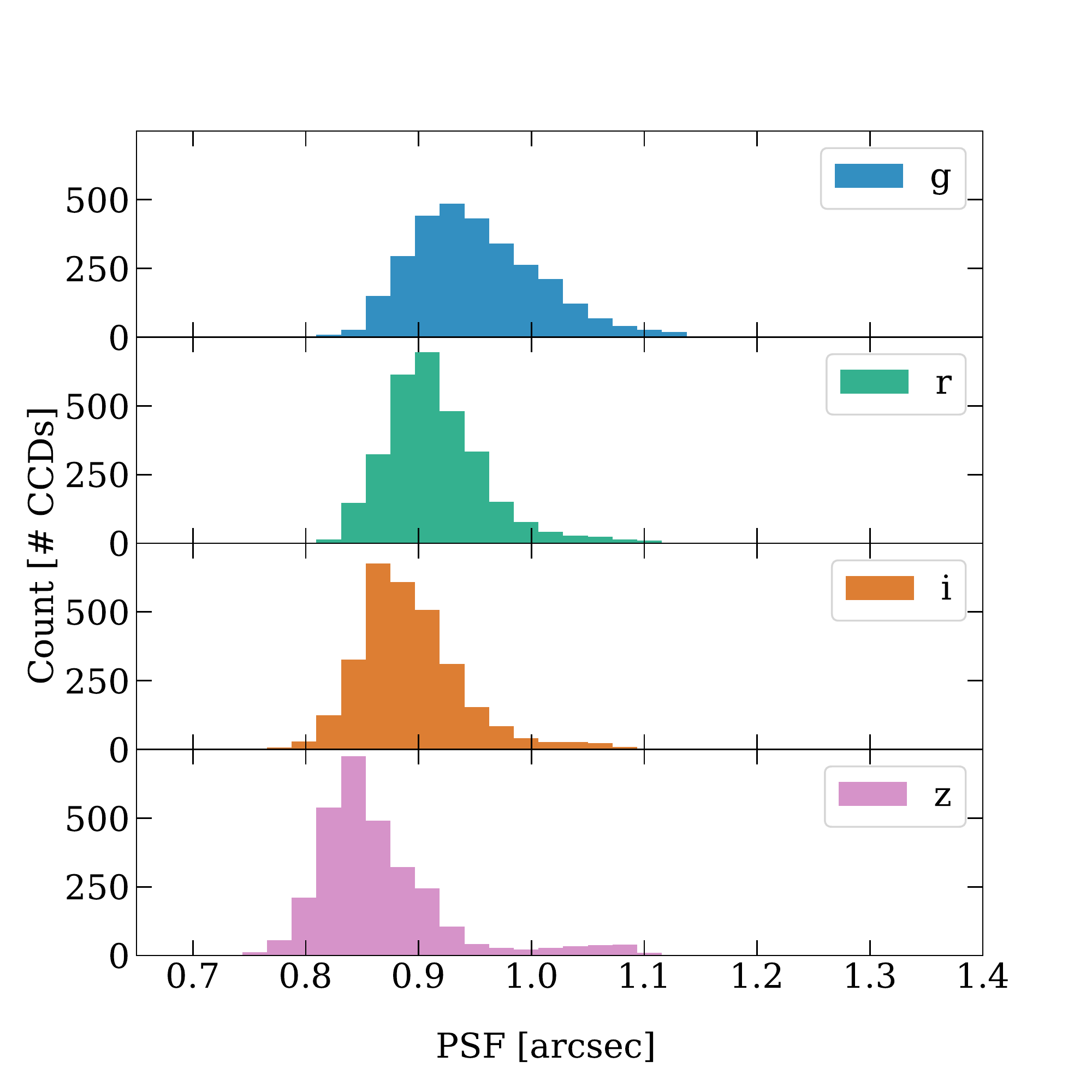}
\caption{Histograms of the average seeing in the seeing-optimised stacks, where each histogram contains the 59 working DECam science CCDs across the 10 DES-SN fields for the 5 years of the SN survey. To see pre-cut input image distributions, please refer to \citetalias{2020MNRAS.495.4040W} Figure 2.}
\label{fig:aveseeing}
\end{center}
\end{figure}

The depth-optimised stacks of \citetalias{2020MNRAS.495.4040W} were optimised for angular resolution by imposing thresholds on the input images, using $\tau$ (the ratio between effective exposure time and the true exposure time due to conditions; see \citetalias{2020MNRAS.495.4040W} Section 2.2.1 and \citealt{Neilsen2016,2019arXiv191206254N}) and the PSF FWHM (hereafter PSF). Input images must pass the cuts (i.e. have a $\tau$ above and PSF below the given threshold). Here, we use the same technique, but optimise the final seeing rather than depth. We tested various combinations of $\tau_\mathrm{cut}$ and PSF$_\mathrm{cut}$. For the seeing-optimised stacks, we find that the $\tau_\mathrm{cut}$ is relatively unimportant, and use the minimum limiting value of $\tau_\mathrm{cut} = 0.02$ to remove clear outliers in image quality. The maximum PSF (PSF$_\mathrm{cut}$) was set at 1.3\arcsec\ in all filters, providing a balance between depth and image quality (and hence redshift coverage). Fig.~\ref{fig:aveseeing} displays the average seeing in each band for the seeing-optimised stacks. 

Full details of the selection cuts are in Table~\ref{table:stack_params}. Our PSF cut is tighter than \citetalias{2020MNRAS.495.4040W}, cutting at 1.3\arcsec\ in all filters, compared to their 2.4\arcsec\ in $g$ band and 2.2\arcsec\ in other bands. However, the \citetalias{2020MNRAS.495.4040W} stacks are deeper, with limiting magnitudes of $\sim26$\,mag; our seeing-optimised stacks have limiting magnitudes of $\sim25$\,mag.  

\begin{table}
\begin{center}
\caption{Seeing-optimised image stack parameters.}% We use $\tau_\textrm{cut}$ = 0.02, and $\textrm{PSF}_\textrm{cut}$ = \SI{1.3}{\arcsecond}.}
\begin{threeparttable}
\begin{tabular}{lllllr}
\hline
 Field\tnote{1} & Band\tnote{2} & MY\tnote{3} & $N_{\textrm{exp}}$\tnote{4} & $t_{\textrm{exp;tot}}$\tnote{5} &  $m_{\textrm{lim}}$\tnote{6} \\
\hline

 SN-E1 &    g &  1  &                  7 &                   0.34 &               26.25 \\
 SN-E1 &    g &  2  &                  3 &                   0.15 &               25.06 \\
 SN-E1 &    g &  3  &                  6 &                   0.29 &               26.24 \\
 SN-E1 &    r &  1  &                 22 &                   0.92 &               25.86 \\
 SN-E1 &    r &  2  &                 14 &                   0.58 &               25.71 \\

\hline
\multicolumn{6}{r}{{Full table available online}} \\
\hline

\end{tabular}
\begin{tablenotes}
\item[1] SN field.
\item[2] Filter band.
\item[3] 'Minus Year' missing season, subtracted to remove contamination from SN light.
\item[4] Number of single exposures in each coadd.
\item[5] Total exposure time given in hours.
\item[6] Limiting magnitude determined from the sky background.
\end{tablenotes}
\end{threeparttable}
\label{table:stack_params}
\end{center}
\end{table}

\subsubsection{Global photometry}
\label{global_phot} 

Following \citetalias{2020MNRAS.494.4426S} and \citetalias{2020MNRAS.495.4040W}, the global photometry for the host galaxy is measured using \textsc{Source Extractor} \citep{1996A&AS..117..393B} on the stacked images. We use \textit{griz} Kron \texttt{FLUX\_AUTO} measurements, using a detection image to set the aperture so that the aperture is the same in arcseconds for the measurement in each filter, and correct for Milky Way dust extinction using \citet{1998ApJ...500..525S} dust maps and Fitzpatrick reddening law \citep{1999PASP..111...63F} with multiplicative coefficients from the first DES data release \citep{2018ApJS..239...18A}: $R_{g} = 3.186, R_{r} = 2.140, R_{i} = 1.569, R_{z} = 1.196$.

\subsubsection{Local photometry} \label{local_phot}

The smaller the aperture, the more representative the photometry is of the stellar population local to the SN site; however, this is limited by the combined effects of the atmosphere and telescope (PSF size) and the coadding procedure on the final images. Assuming a maximum FWHM of 1.3\arcsec\ (Section~\ref{stacks}), and a Gaussian PSF with $\mathrm{FWHM} = 2\sqrt{2 \mathrm{ln} 2} \approx 2.355\sigma$, we have a smallest useful aperture radius ($\sigma$) of 0.55\arcsec. Coincidentally, this is approximately the DECam corrector's contribution to the PSF, i.e. the best PSF that can be achieved in near perfect sky conditions.

\begin{figure}
\includegraphics[width=\linewidth]{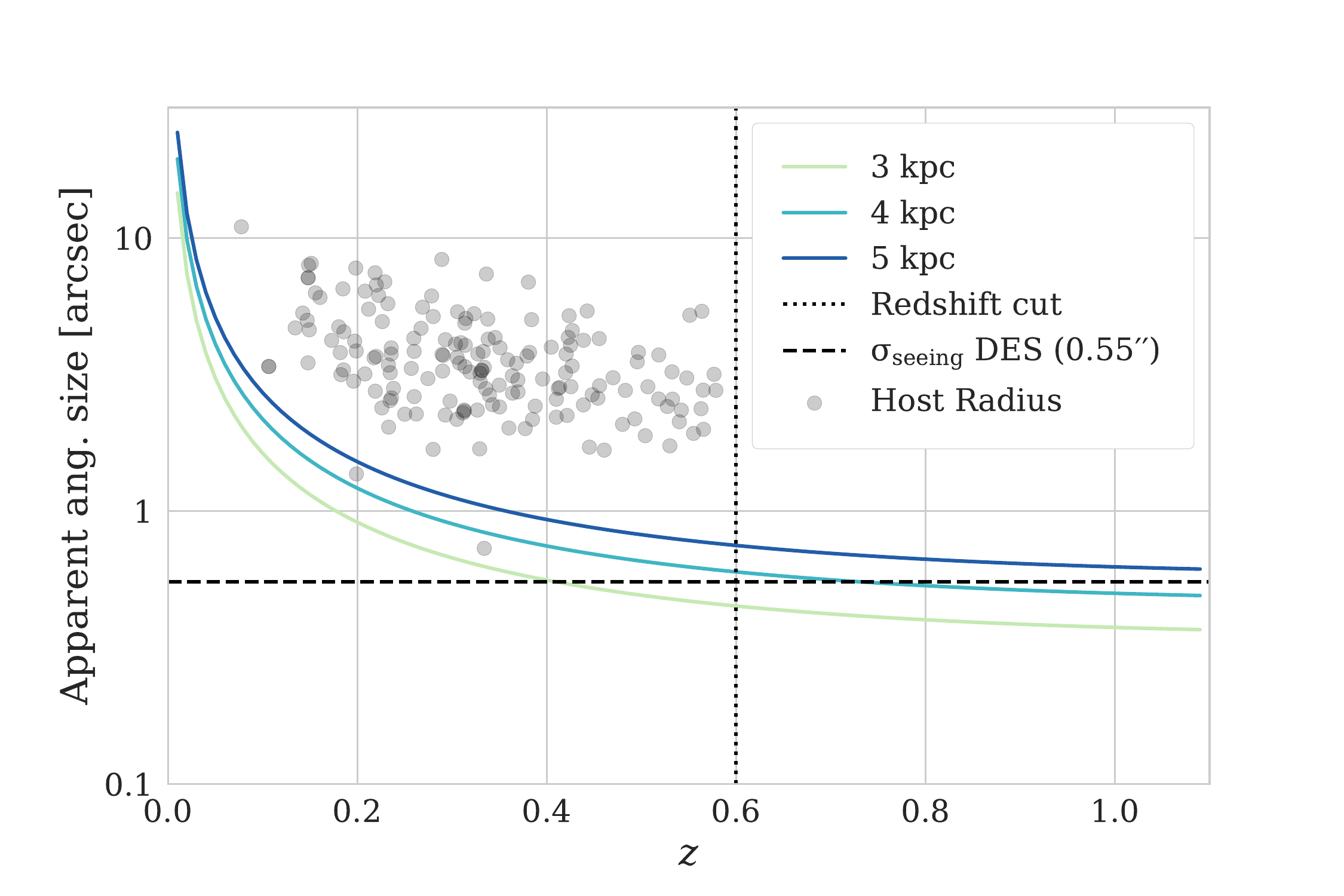}
\caption{The evolution of the apparent angular size in arcseconds with redshift, for 3\,kpc, 4\,kpc and 5\,kpc local aperture radii. The dashed horizontal line indicates the $1\sigma$ seeing of the DES seeing-optimised stacks, and the dotted line the $z<0.6$ redshift cut. For comparison, the grey shaded circles are representative of the radius (in arcseconds) of a circle with equal area to that of the \textsc{Source Extractor} detected ellipse for each host galaxy used in our analysis.}
\label{fig:ap_z}
\end{figure}

\begin{figure*}

    \includegraphics[width=5.8cm]{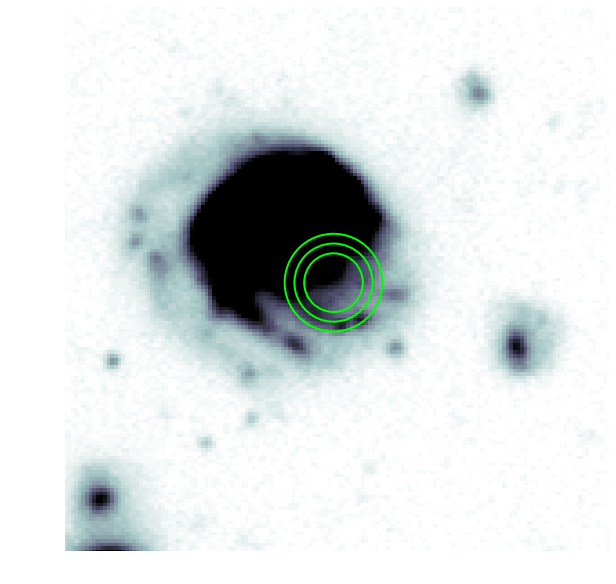}
    \includegraphics[width=5.8cm]{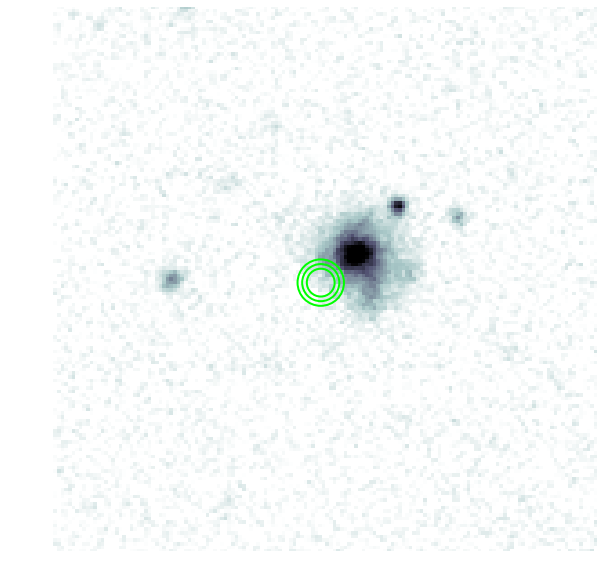}
    \includegraphics[width=5.8cm]{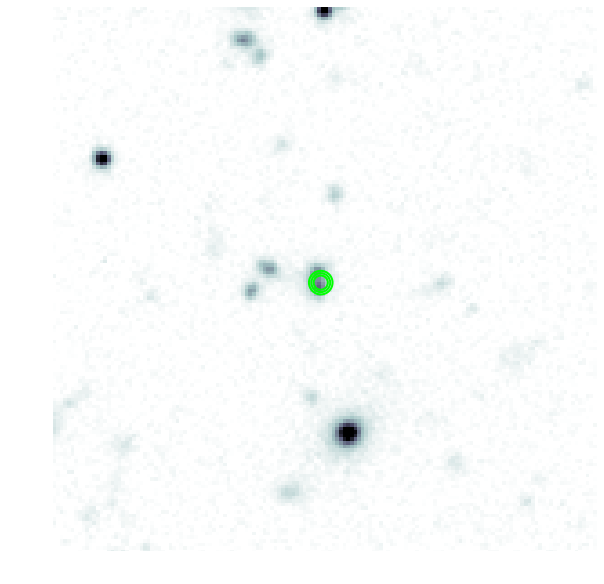}
    \hspace{3.0cm}
    \subfloat{DES15C3efn : $z = 0.078$}
    \hspace{3.0cm}
    \subfloat{DES13E1sae : $z = 0.185$}
    \hspace{3.0cm}
    \subfloat{DES15C3nym : $z = 0.496$}
    \hspace{3.0cm}

\caption{Three $g$-band images of DES SN Ia host galaxies at $z<0.6$. The green circles represent the local region within a 3\,kpc, 4\,kpc and 5\,kpc aperture radius centred on the SN location. All images are set with the same image intensity scaling parameters, and are of the same angular scale.}

\label{fig:ap}
\end{figure*}

This motivates the common physical aperture size we apply in our measurements. In Fig.~\ref{fig:ap_z}, we show the apparent size of 3\,kpc, 4\,kpc and 5\,kpc physical apertures as a function of redshift. At about $z=0.7$, the 4\,kpc aperture becomes smaller than a 0.55\arcsec\ radius, and thus to be conservative we safely select $z=0.6$ as a redshift cut that we apply to all our DES SNe Ia, together with a consistent 4\,kpc radius aperture for our analysis. We note that such a redshift cut also minimises selection bias on our sample, particularly in the shallow fields \citep[1D $\mu_\mathrm{bias} \sim-0.06$\,mag at $z=0.6$; for 5D, see][]{2019MNRAS.485.1171K}. We discuss the affect of varying this aperture size in Section~\ref{varying}. Examples of the aperture regions probed in relation to galaxy size and redshift are shown in Fig.~\ref{fig:ap}. 

We perform the local aperture photometry using \textsc{aperture\_photometry} tool from the \textsc{photutils} Python module \citep{larry_bradley_2019_2533376}. Photometric uncertainties are calculated using the weight maps associated with each stack. We correct the resulting fluxes for Milky Way extinction in the same way as the global photometry. 

\subsection{SED fitting} \label{SED}

In our analysis, we use spectral energy distribution (SED) fitting techniques for both global and local host galaxy properties, fitting galaxy templates to our photometry. We use the same templates for fitting both the local and global photometry, i.e., in essence, we treat each local region as a small galaxy. 

We estimate the environmental parameters of the host galaxies and local regions following \citetalias{2020MNRAS.494.4426S} and references therein. Our SED fitting and templates are based on the \textsc{p\'egase} spectral evolution code \citep{1997A&A...326..950F, 2019A&A...623A.143F}. We assume a  \citet{2001MNRAS.322..231K} initial mass function (IMF) and a series of 9 smooth exponentially-declining star-formation histories, with 102 time steps in each. We generate synthetic DES $griz$ photometry for each SED and compare with the observed $griz$ photometry via a standard $\chi^2$ minimisation. All fitting is done in flux space, and we only consider solutions younger than the age of the universe at each SN redshift. We also consider foreground dust screens with a colour excess $E(B-V) = 0$ to $0.3$\,mag in steps of $0.05$\,mag. 

This fitting determines the environmental properties of either the global host galaxy or the local region: the star formation rate (SFR, in $\mathrm{M}_{\sun}$yr$^{-1}$, averaged over the last 0.25\,Gyr before the best-fitting time step), \mstellar, and the sSFR (in yr$^{-1}$). To estimate the statistical uncertainties in these parameters, we use a Monte Carlo process adjusting the observed photometry according to its uncertainties, with 1000 iterations for each host galaxy (or local region). 

We estimate the rest-frame $UBVR$ magnitudes by taking the best-fit SED for each SN Ia host galaxy fit for each random realisation in the Monte Carlo, and adjusting that SED using a wavelength-dependent multiplicative function so that the SED exactly reproduces the observed $griz$ photometry, a process sometimes referred to as \lq mangling\rq\ \citep{2007ApJ...663.1187H, 2008ApJ...681..482C}. We use a spline function as the multiplicative function, and follow the same Monte Carlo process to estimate the statistical uncertainty.

In this paper, we focus our analysis on the rest-frame $U-R$ colour, as this spans the greatest wavelength range covered by our observer-frame ($griz$) photometry. $U-R$ correlates with galaxy morphology \citep[as seen in the correlation with $u-r$;][]{2008MNRAS.389.1179L}, is a complementary tracer of the SFR, and carries information about the age of the SN host galaxy. This relationship is due to the different filter responses being dominated by different types of emission: the older stars or more passive galaxies at the redder end of the spectrum, and the younger, hotter stars or more star-forming galaxies at the bluer end \citep{2016MNRAS.460.3925T}.

\subsection{Selection requirements} \label{selection_cuts}

We make two additional cuts to the 206 SNe Ia from the DES3YR cosmological sample. Firstly, as motivated in Section~\ref{local_phot}, we require $z<0.6$ to obtain relevant local photometry. Secondly, we require the SN hosts to have well-measured rest-frame $U-R$ colour and therefore require the $U-R$ uncertainty: $\sigma_{(U-R)} < 1$\,mag for both the global and local measurements. As the $U-R$ value is derived from the observed photometry, this cut also removes those events with large uncertainties in \mstellar\ and SFR.

\begin{figure}
\begin{center}
\includegraphics[width=\linewidth]{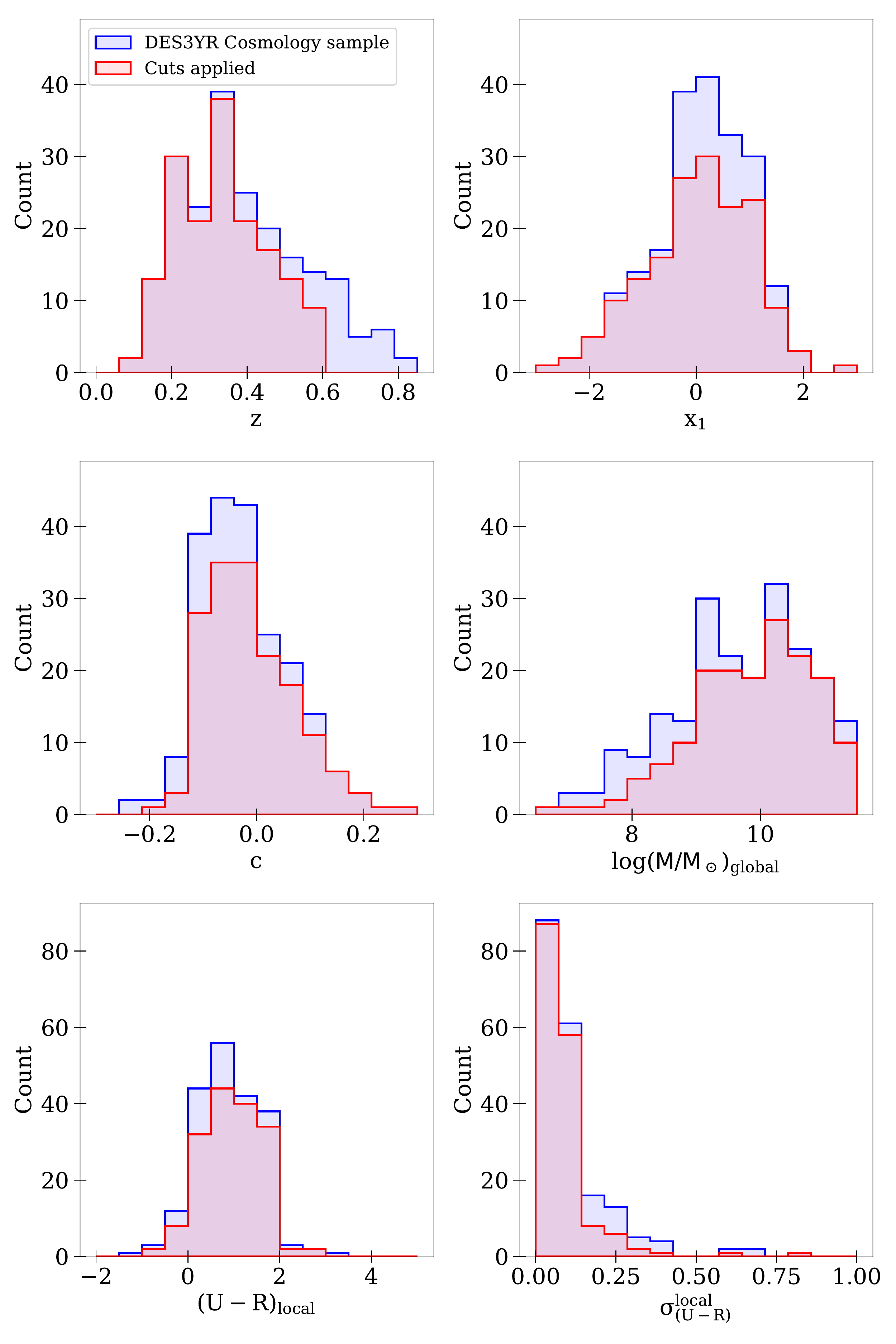}
\caption{Histograms of the distributions of redshift ($z$), SN stretch ($x_1$), SN colour ($c$), host \mstellar, local rest-frame $U-R$ colour (in a 4\,kpc aperture radius), and local $U-R$ colour uncertainty ($\sigma_{(U-R)}^{\rm local} < 1$). The blue-shaded histogram represents the entire DES3YR cosmology sample, and the red histogram is after cuts in Table~\ref{table:selection}.
}
\label{fig:4hists}
\end{center}
\end{figure}

After cuts, 164 objects remain in our sample (Table~\ref{table:selection}). Fig~\ref{fig:4hists} demonstrates that our selection cuts have only minor effects on the distributions of $x_1$, $c$, \mstellar, and the local $U-R$ colour. 

\begin{table}
\begin{center}
\caption{Sample selection cuts used for our analysis.}
\begin{tabular}{c c} 
\hline
Cut & Number of SNe Ia\\
\hline
Cosmology Sample & 206 \\
Redshift Cut  & 177 \\ 
$\sigma_{(U-R)} < 1$ & 164 \\
\hline
\end{tabular}
\label{table:selection}
\end{center}
\end{table}

\section{Environmental dependence of SN Ia luminosities} \label{results}

Having measured global and local photometry of the SN Ia host galaxies and inferred various physical properties of the stellar populations, we now turn to analysing such data in SNIa standardisation for cosmological analyses.

\subsection{Global vs. Local measurements} 

\begin{figure*}
\begin{center}
\includegraphics[width=.45\linewidth]{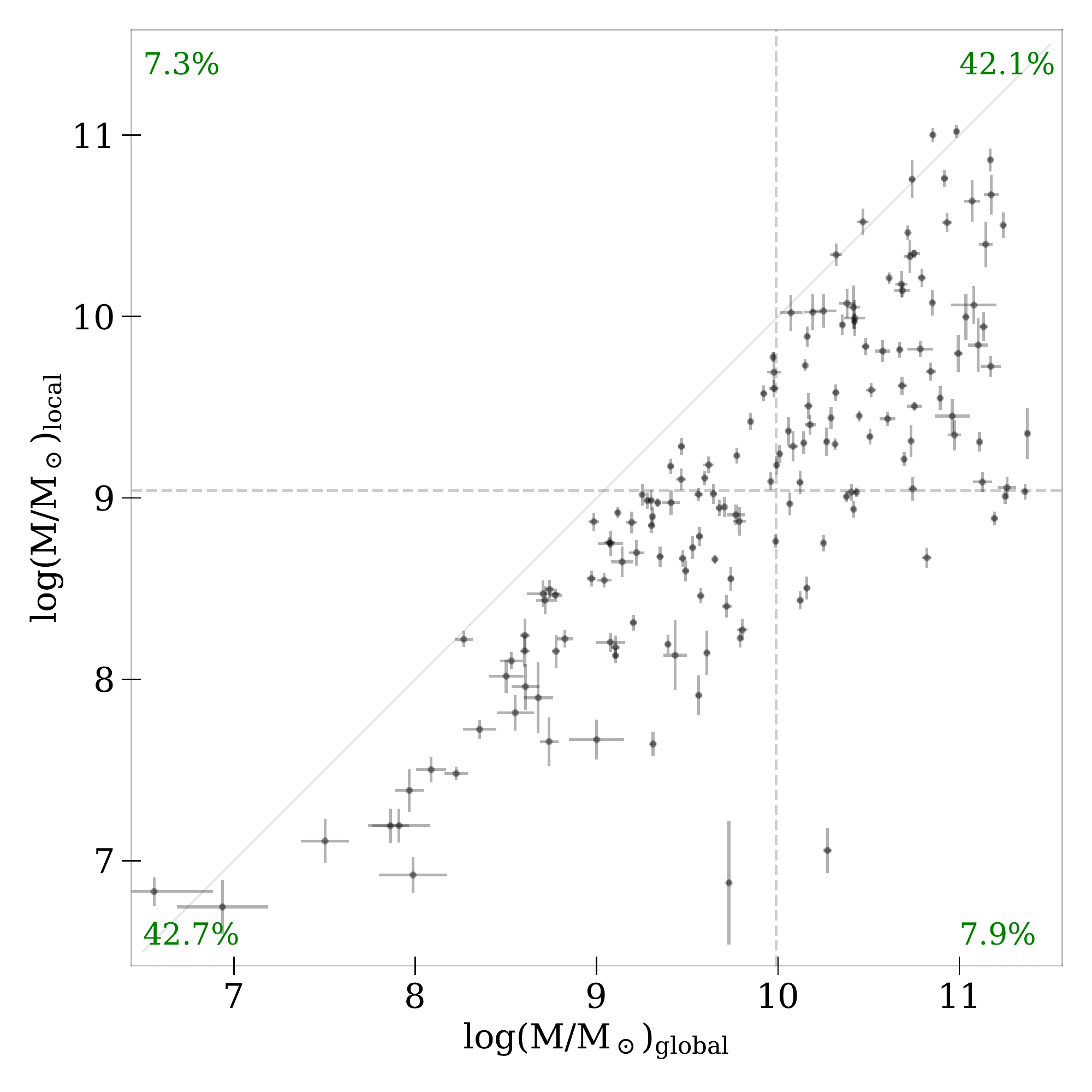}
\includegraphics[width=.45\linewidth]{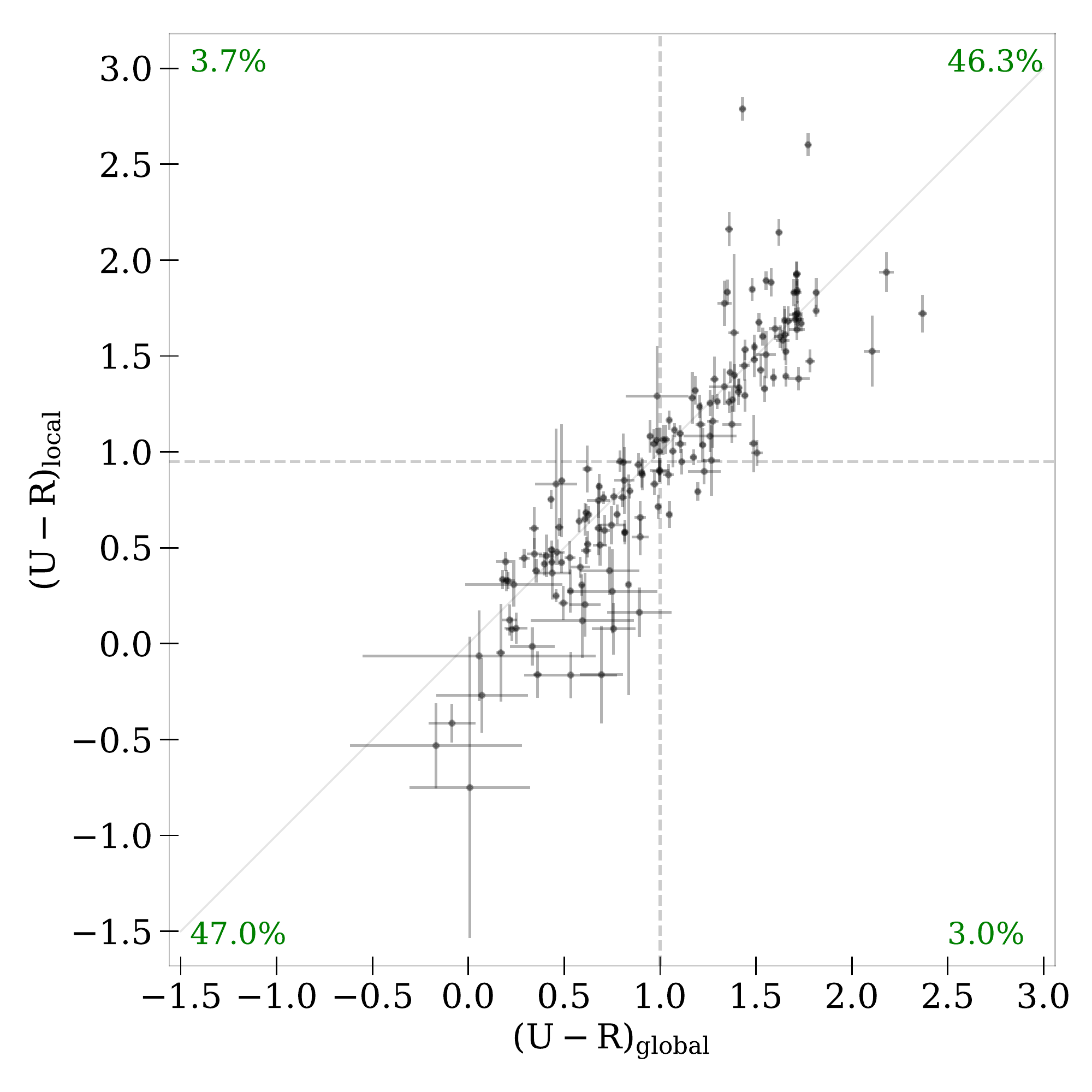}
\vspace{-10mm}
\includegraphics[width=.45\linewidth]{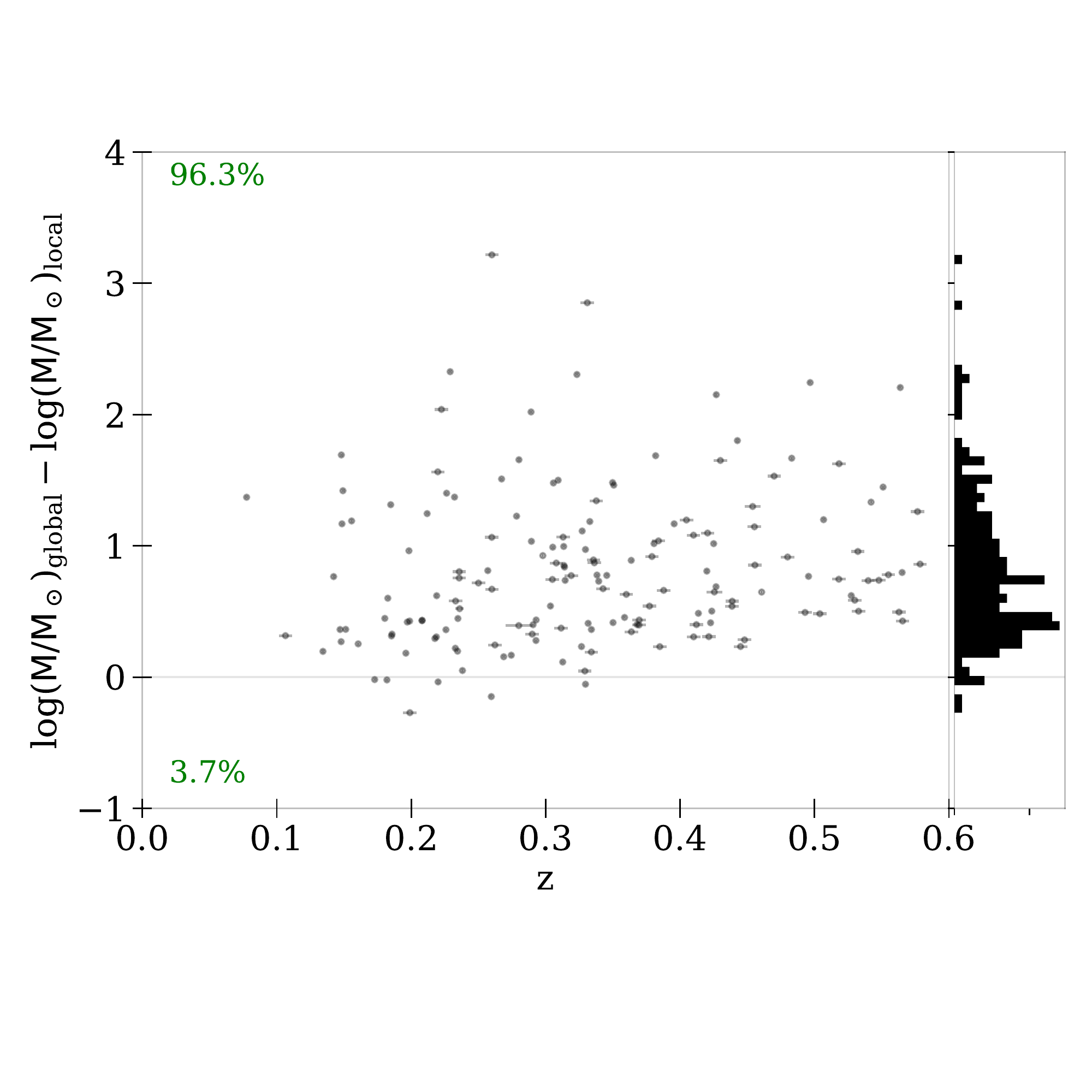}
\includegraphics[width=.45\linewidth]{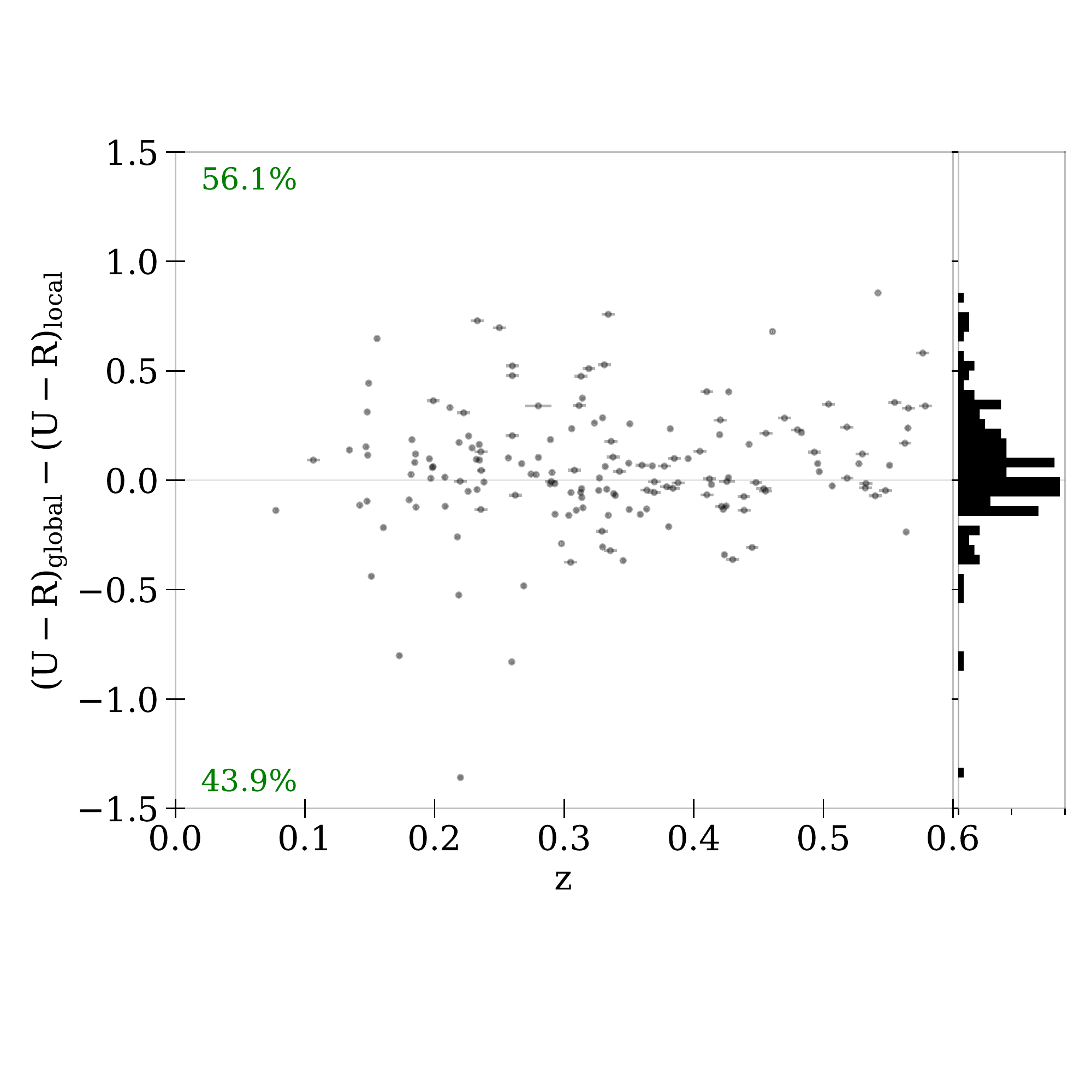}
\caption{Left: The difference between the global stellar mass of the host galaxy and the local stellar mass in the 4\,kpc radius aperture around the SN location. Right: The difference between the global rest-frame $U-R$ colour and the local rest-frame $U-R$ colour. The solid line shows the 1:1 line (matching the zero difference line in the lower panel), dashed lines indicate the environmental property median points of the sample, and green percentages represent the numbers of agreement in each quadrant (e.g., what percentage of the sample are both high local colour and high global colour, etc.). For stellar mass the Pearson correlation coefficient, $r = 0.801$, and for rest-frame $U-R$: $r = 0.895$. In the lower panels, the difference in properties versus redshift is shown. Green percentages represent the proportions of the sample above and below the zero difference line. Error bars throughout represent the statistical uncertainty in the data.}
\label{fig:z_vs_diff}
\end{center}
\end{figure*}

We first compare the global and local properties of the SN Ia host galaxy sample (Fig.~\ref{fig:z_vs_diff}), and the \lq global minus local\rq\ differences as a function of redshift. As expected, the local regions typically have smaller stellar mass values than the global stellar mass, with no strong trend with redshift. There are a few SN hosts with a higher local stellar mass, but these have a large uncertainty or may represent those where the aperture is probing a region larger than the host galaxy.

Both comparisons have statistically significant scatter, indicating that local and global measurements provide different information reflecting the local stellar populations, with the scatter slightly larger for stellar mass than for $U-R$. The $U-R$ colour difference is slightly positive, indicating that SNe Ia have a slight preference for bluer, presumably stronger star-forming local environments than their host-galaxy average. This preference for bluer regions is consistent with earlier studies \citep{2015MNRAS.448..732A}.

\subsection{SN properties vs. environments} \label{prop_v_host}

\begin{figure*}
\begin{center}
\vspace{-4mm}
\includegraphics[width=.45\linewidth]{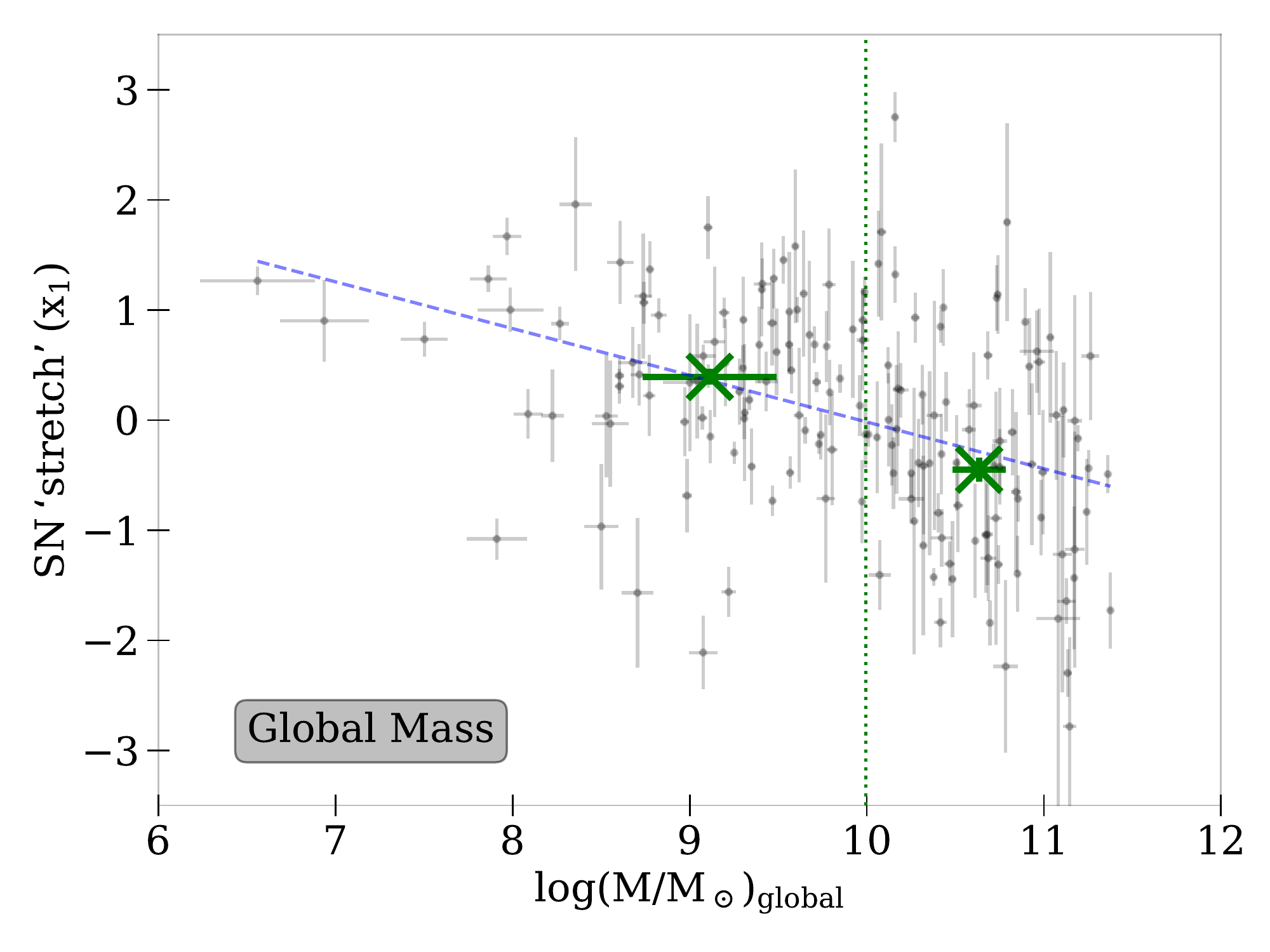}
\includegraphics[width=.45\linewidth]{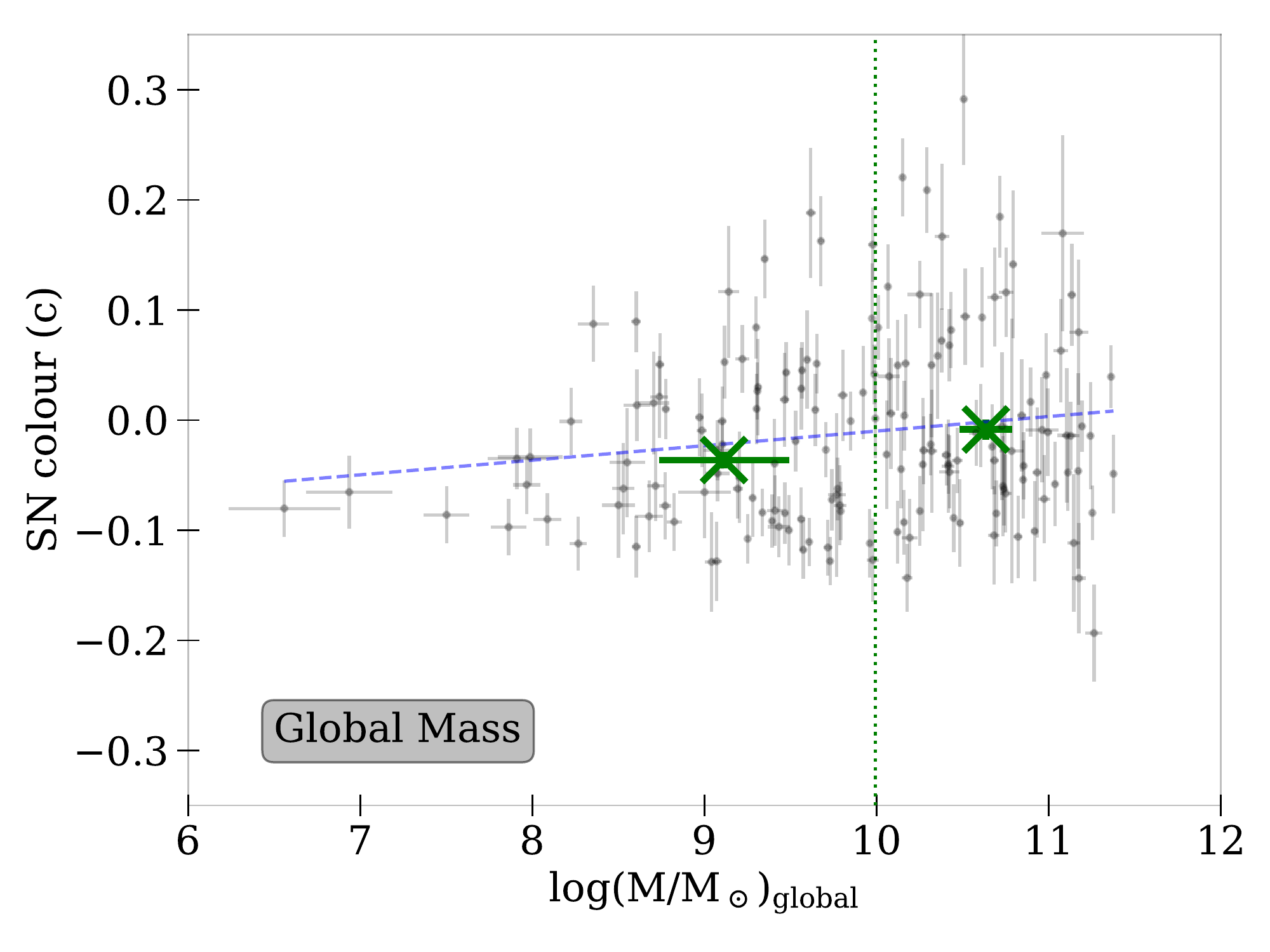}
\vspace{-2mm}
\includegraphics[width=.45\linewidth]{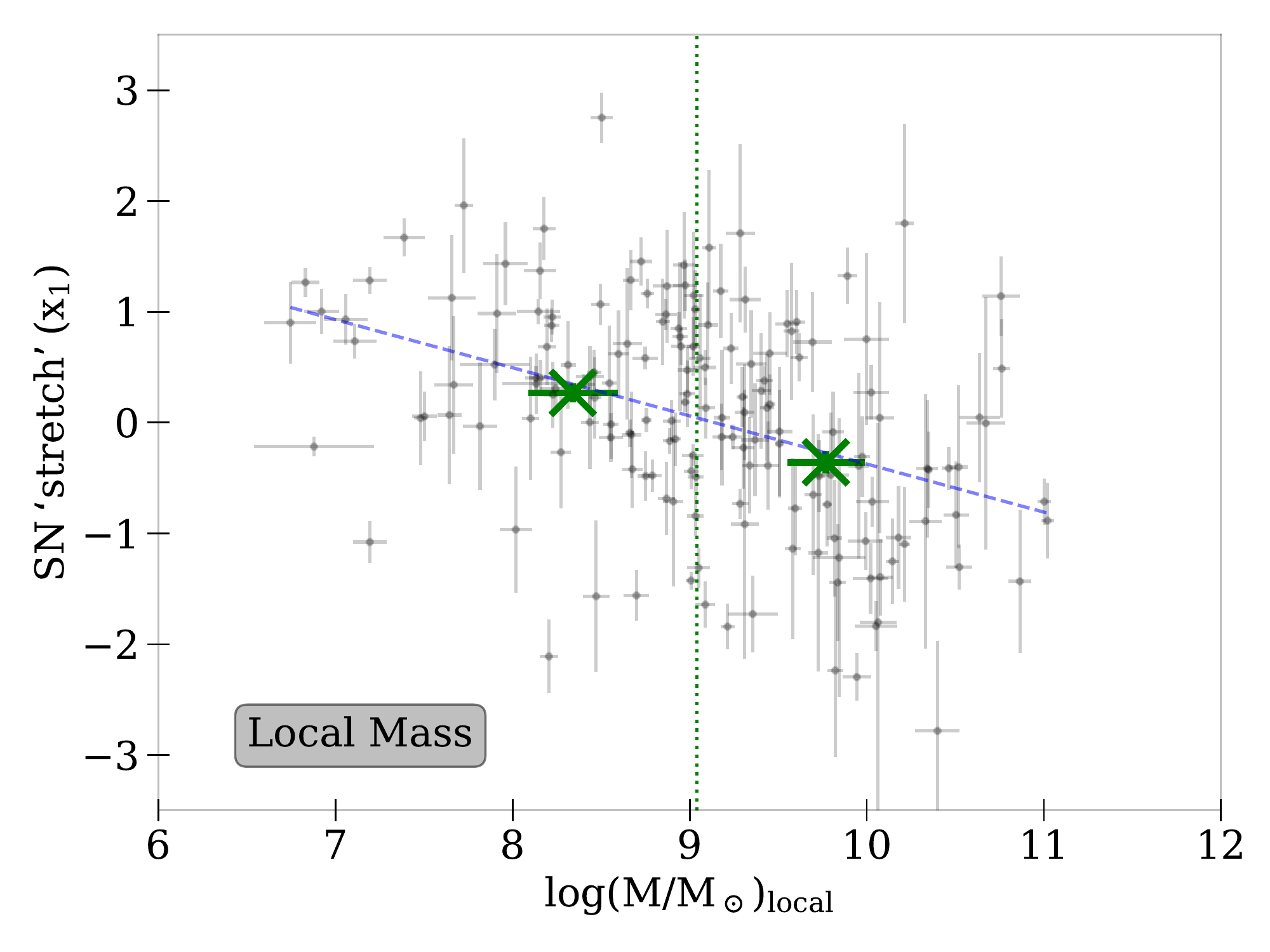}
\includegraphics[width=.45\linewidth]{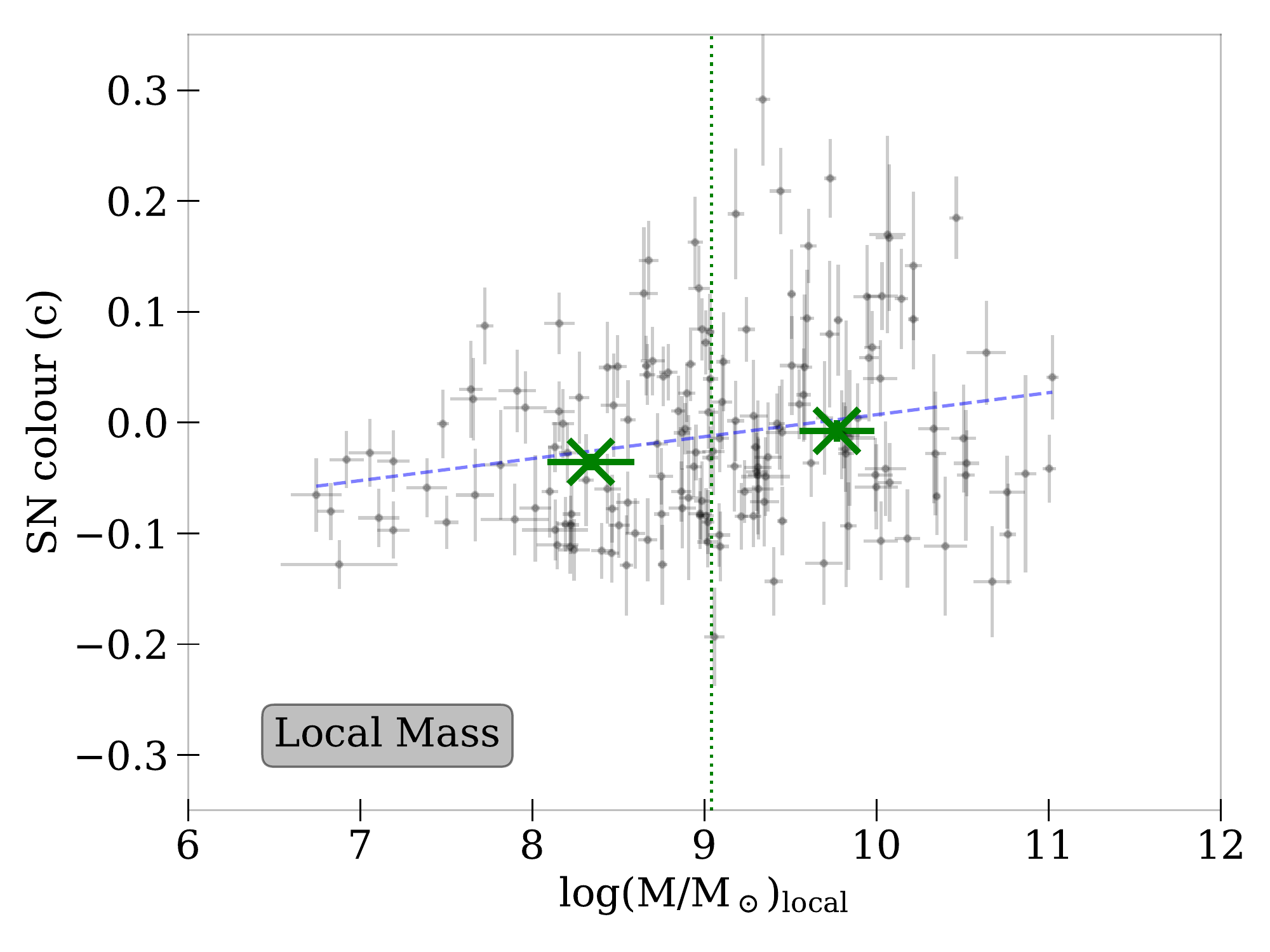}
\vspace{-2mm}
\includegraphics[width=.45\linewidth]{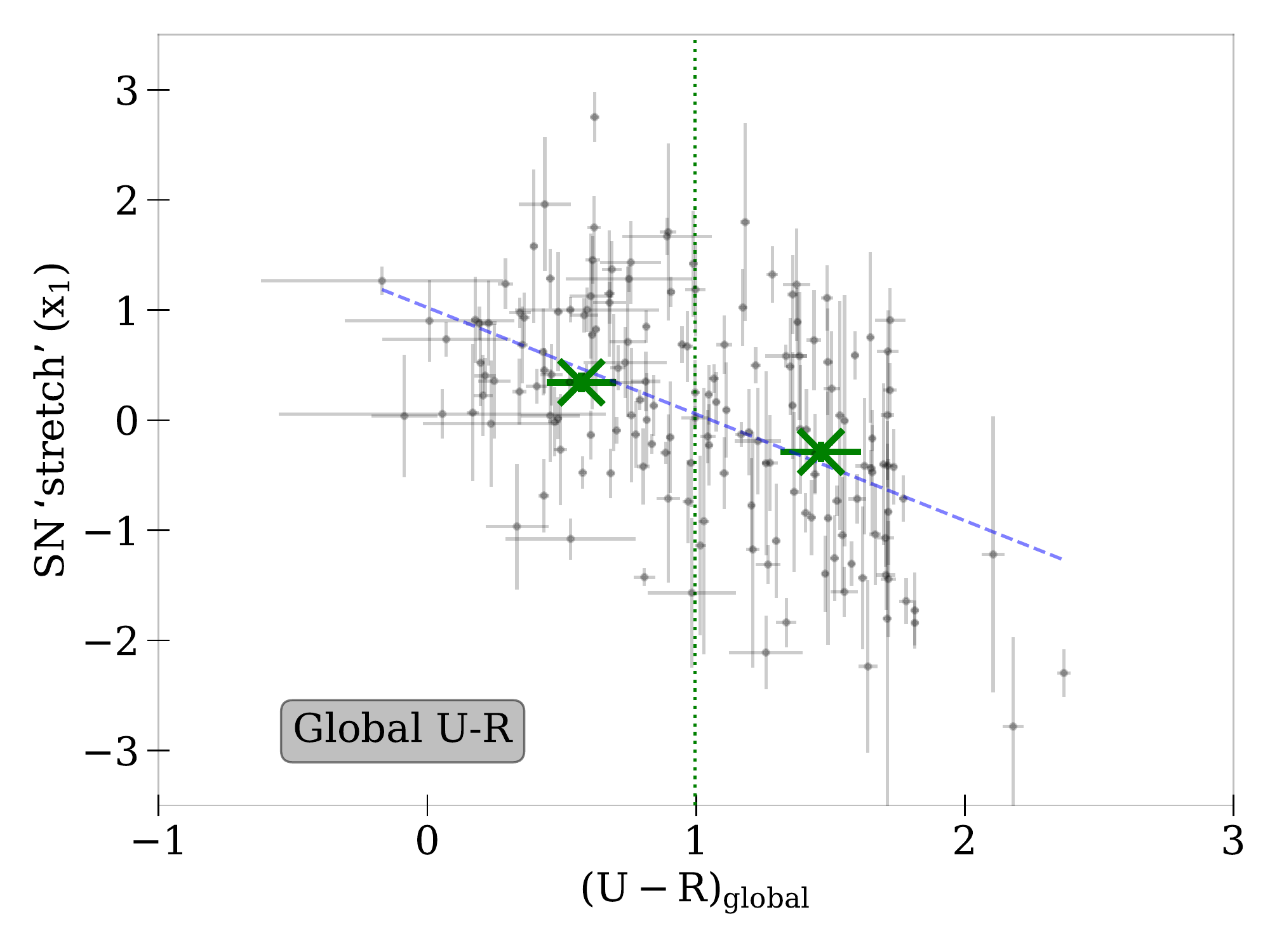}
\includegraphics[width=.45\linewidth]{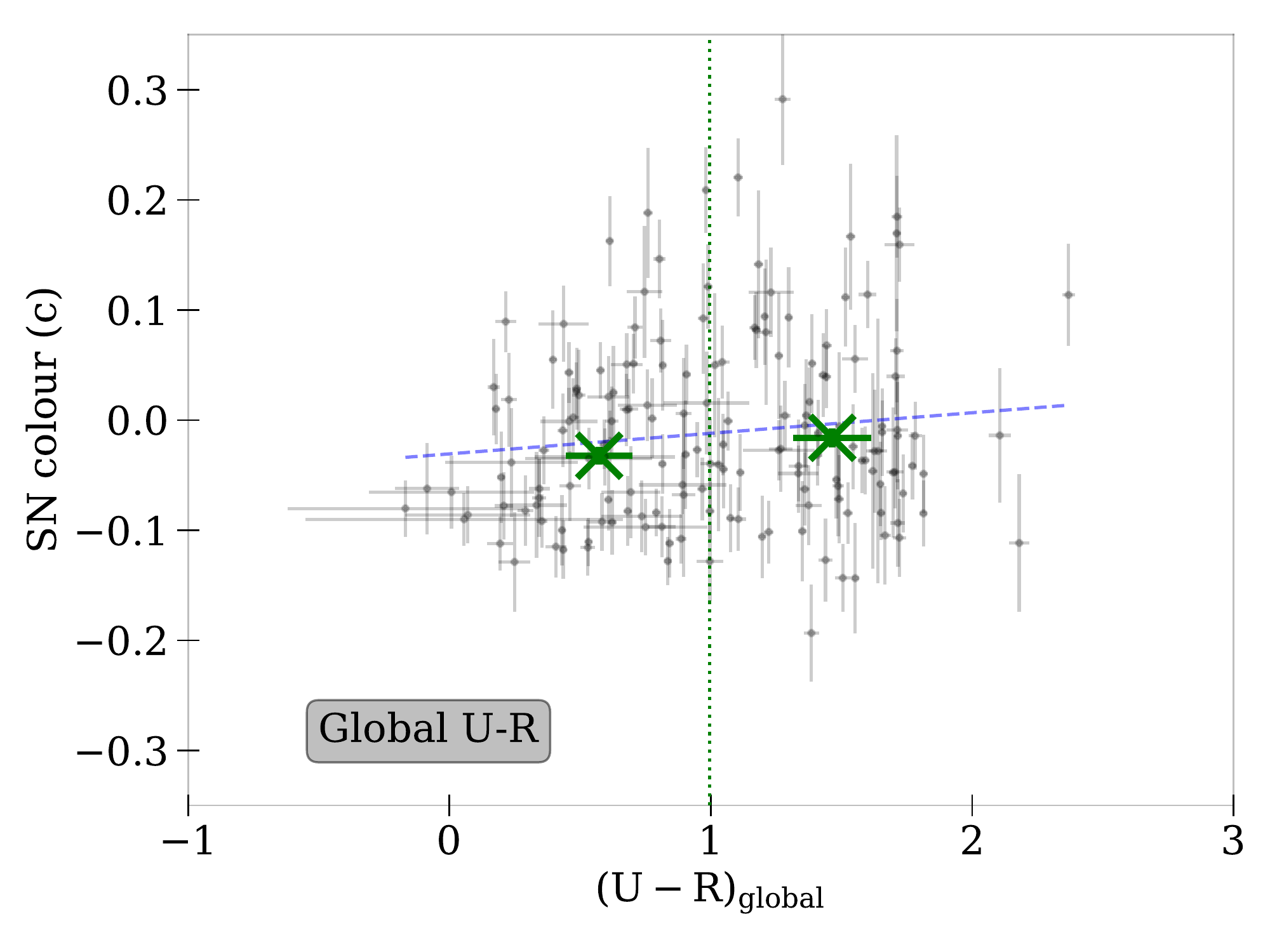}
\vspace{-4mm}
\includegraphics[width=.45\linewidth]{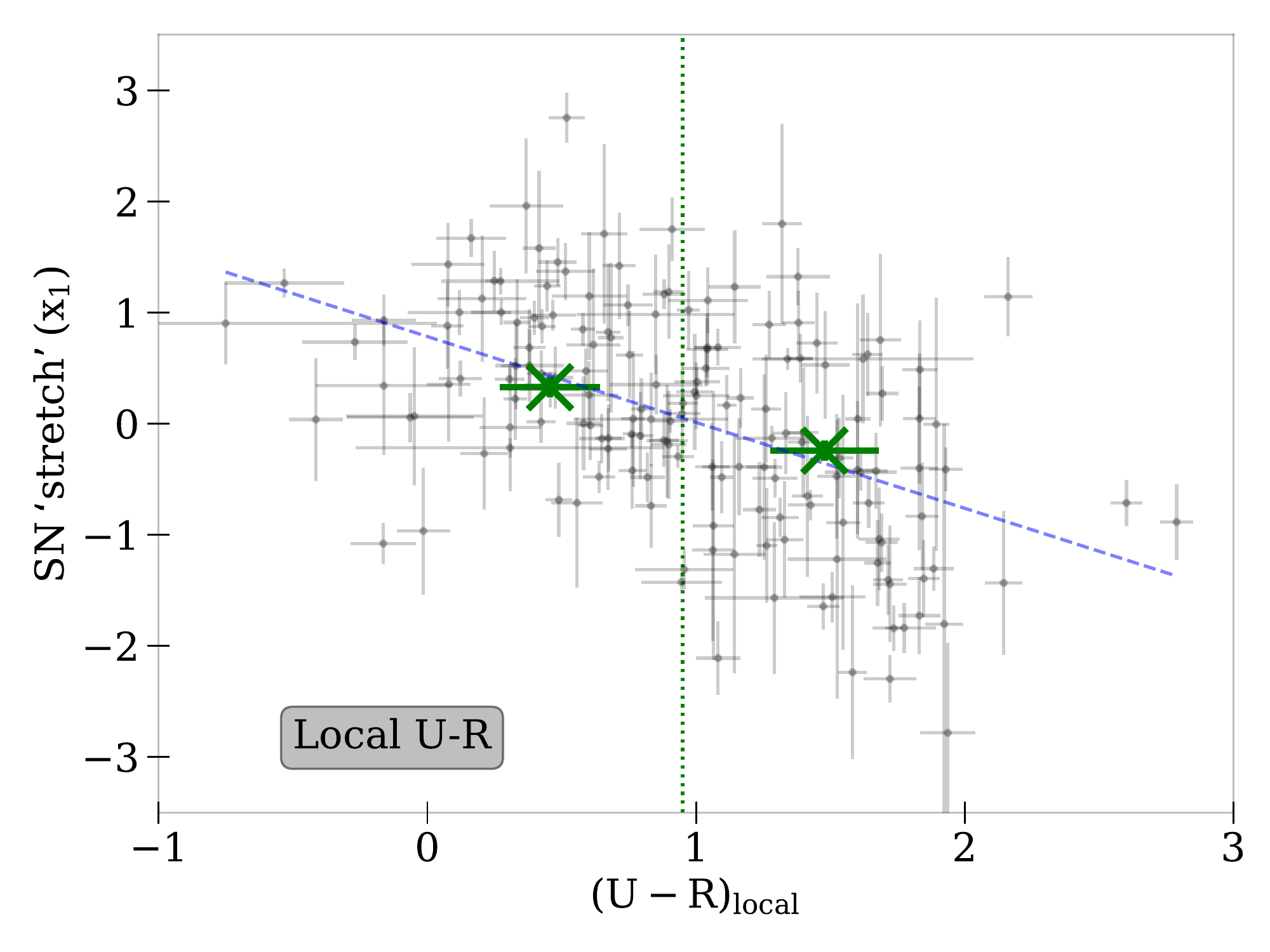}
\includegraphics[width=.45\linewidth]{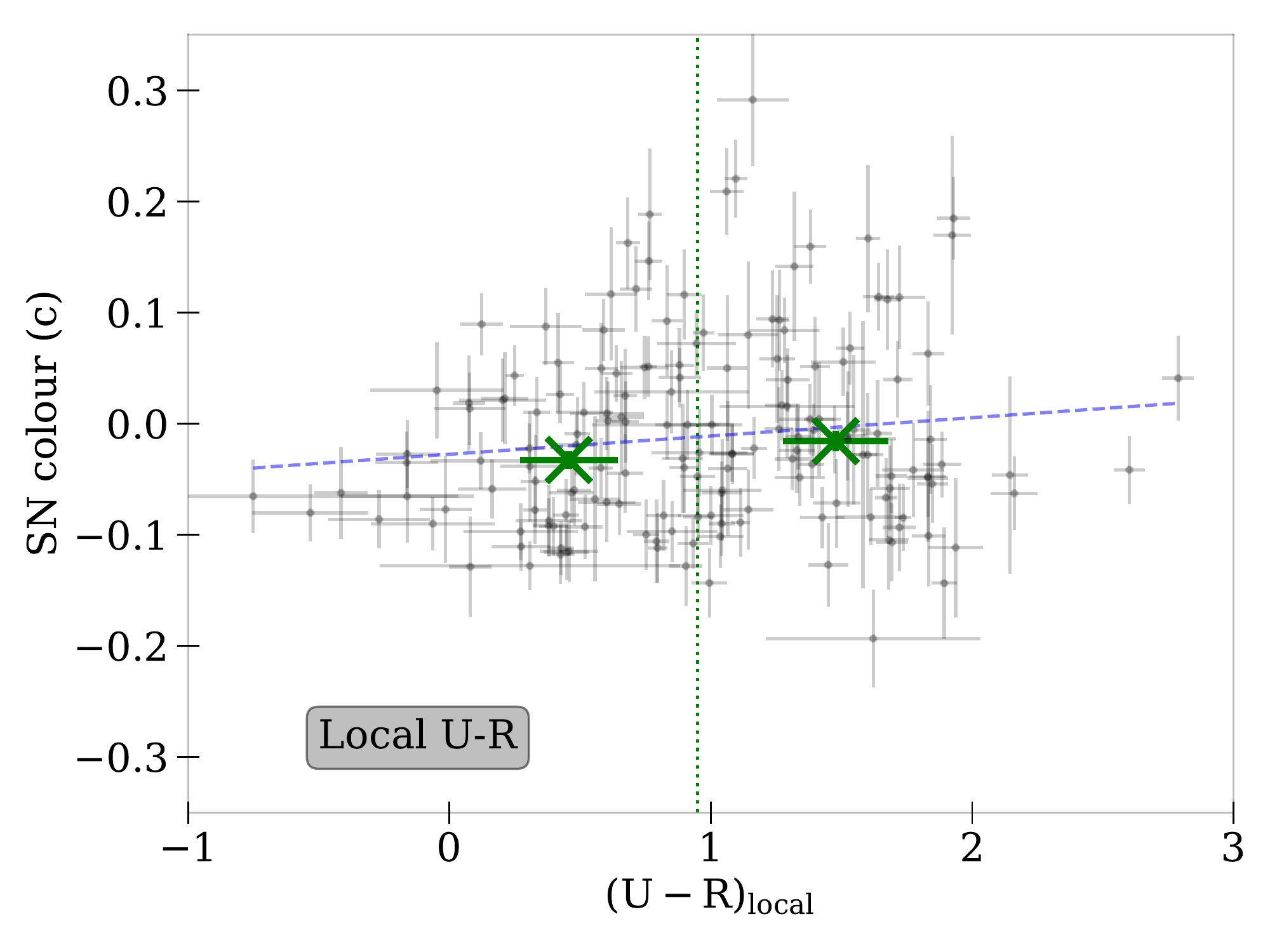}

\caption{Correlations between SN stretch ($x_1$, left column) and colour ($c$, right column) as a function of: global \mstellar\ (first row), local \mstellar\ within a 4\,kpc radius aperture (second row), global $U-R$ (third row) and local $U-R$ (fourth row). Bins are split at the median of the sample, with weighted mean values shown as crosses, x-axis bin-mean error bars showing the dispersion divided by the square root of the number of objects in the bin, and least squares linear fits of the data shown as dashed lines to aid the eye. Corresponding r.m.s. values can be found in Table~\ref{table:4values_x1_c}.}
\label{fig:prop_comp} 
\end{center}
\end{figure*}

In Fig.~\ref{fig:prop_comp} and Table~\ref{table:4values_x1_c} we show the relationship between SN $x_1$ and SN $c$, and the rest-frame $U-R$ colour and \mstellar\ of the global host galaxy and the local SN environment. Strong trends are evident in the $x_1$ comparisons, with brighter-slower SNe Ia in bluer, less massive environments, and mild trends in the $c$ comparisons, with redder SNe Ia in more massive galaxies. Additionally, as can be seen in the top panel corresponding to global stellar mass, there is an absence of fast evolving and red SNe in low-mass galaxies. 

We also divide the sample into bins of $U-R$ and \mstellar\ (both global and local), and calculate the difference in the mean $x_1$ and $c$, as well as the r.m.s. of each sample (in Table~\ref{table:4values_x1_c}). We find that the $x_1$ difference is most significant for the global \mstellar, recovering the known relationship between $x_1$ and \mstellar. 

Table~\ref{table:4values_x1_c} also shows that SNe Ia in more massive galaxies or environments have a higher r.m.s in the SN $x_1$ and $c$ populations. This is also the case for the redder environments (larger $U-R$ values). SNe Ia in the more star-forming, bluer regions present a more homogeneous sample.

\begin{table*}
\begin{center}
\caption{Stretch ($x_1$) and colour ($c$) variation with host galaxy stellar mass and $U-R$ colour (Fig.~\ref{fig:prop_comp}).}
\begin{threeparttable}
\resizebox{\textwidth}{!}{
\begin{tabular}{c c c c c c c c c c}

\hline 
\multicolumn{1}{c}{Property} & \multicolumn{1}{c}{Division}  & \multicolumn{2}{c}{$x_1$} & \multicolumn{2}{c}{$x_1$ RMS} & \multicolumn{2}{c}{$c$} & \multicolumn{2}{c}{$c$ RMS}\\ & \multicolumn{1}{c}{Point\tnote{1}} & Sig. ($\sigma$)\tnote{2} & Magnitude\tnote{3} & < DP\tnote{4} & > DP & Sig. ($\sigma$) & Magnitude & < DP & > DP\\
\hline
Global Mass & 9.99 & 6.50 & 0.842 $\pm$ 0.130 & 0.875 $\pm$ 0.137 & 1.066 $\pm$ 0.169 & 2.32 & 0.028 $\pm$ 0.012 & 0.079 $\pm$ 0.012 & 0.091 $\pm$ 0.014\\ 
Local Mass & 9.04 & 4.71 & 0.626 $\pm$ 0.133 & 0.924 $\pm$ 0.144 & 1.023 $\pm$ 0.162 & 2.30 & 0.028 $\pm$ 0.012 & 0.075 $\pm$ 0.012 & 0.094 $\pm$ 0.015\\ 
\hline
Global U-R & 1.00 & 4.94 & 0.632 $\pm$ 0.128 & 0.918 $\pm$ 0.146 & 1.025 $\pm$ 0.159 & 1.37 & 0.016 $\pm$ 0.012 & 0.080 $\pm$ 0.013 & 0.090 $\pm$ 0.014\\ 
Local U-R & 0.95 & 4.42 & 0.571 $\pm$ 0.129 & 0.893 $\pm$ 0.140 & 1.049 $\pm$ 0.165 & 1.42 & 0.017 $\pm$ 0.012 & 0.078 $\pm$ 0.012 & 0.091 $\pm$ 0.014\\ 
\hline
\end{tabular}
}
\begin{tablenotes}
\item[1] Splitting at the sample median.
\item[2] Significance of the difference in $\sigma$.
\item[3] Magnitude difference.
\item[4] Division Point.
%\item[2] Masses in $\log_{10}(M/M_{\sun})$
%\item[3] Within a 4\,kpc radius
\end{tablenotes}
\end{threeparttable}
\label{table:4values_x1_c}
\end{center}
\end{table*}

Similar relationships were explored in previous work by \cite{2018A&A...615A..68R} for local \mstellar\ and rest-frame $U-V$ colour within a 3\,kpc radius, and we find consistent results. As in previous work, we find a significant dependency of the SN $x_1$ \citep{2009ApJ...691..661H, 2009ApJ...707.1449N, 2006ApJ...648..868S} and $c$ \citep{2010MNRAS.406..782S,2013ApJ...770..108C} on environment. 

\subsection{Hubble Residuals} \label{hubble}

\begin{table*}
\begin{center}
\caption{Hubble residual steps for stellar mass and $U-R$ using a 1D bias correction; shown in Fig.~\ref{fig:hubble_mass_colour_step}.}
\begin{threeparttable}
\begin{tabular}{c c c c c c c}

\hline 
\multicolumn{1}{c}{Property} & \multicolumn{1}{c}{Sample Median/} & \multicolumn{1}{c}{Division} & \multicolumn{2}{c}{Hubble Residual} & \multicolumn{2}{c}{Hubble Residual RMS}\\&  \multicolumn{1}{c}{Max Significance\tnote{1}} & \multicolumn{1}{c}{Point} & Sig. ($\sigma$) & Magnitude & < DP & > DP \\
\hline
Global Mass & Median & 9.99 & 3.25 & 0.057 $\pm$ 0.017 & 0.118 $\pm$ 0.019 & 0.142 $\pm$ 0.022\\ 
Global Mass & Max & 9.73 & 4.14 & 0.070 $\pm$ 0.017 & 0.108 $\pm$ 0.019 & 0.145 $\pm$ 0.021\\ 
Local Mass & Median & 9.04 & 3.57 & 0.064 $\pm$ 0.018 & 0.102 $\pm$ 0.016 & 0.154 $\pm$ 0.024\\ 
Local Mass & Max & 9.28 & 5.47 & 0.098 $\pm$ 0.018 & 0.115 $\pm$ 0.017 & 0.150 $\pm$ 0.026\\ 
\hline
Global U-R & Median & 1.00 & 4.73 & 0.081 $\pm$ 0.017 & 0.110 $\pm$ 0.017 & 0.149 $\pm$ 0.023\\ 
Global U-R & Max & 0.95 & 5.29 & 0.088 $\pm$ 0.017 & 0.110 $\pm$ 0.018 & 0.146 $\pm$ 0.022\\ 
Local U-R & Median & 0.95 & 4.84 & 0.082 $\pm$ 0.017 & 0.109 $\pm$ 0.017 & 0.149 $\pm$ 0.023\\ 
Local U-R & Max & 0.90 & 5.15 & 0.085 $\pm$ 0.017 & 0.111 $\pm$ 0.018 & 0.146 $\pm$ 0.022\\
\hline
\end{tabular}
\begin{tablenotes}
\item[1] The difference between median locations/max significance locations is explained in Section~\ref{hubble} and in Figure~\ref{fig:sig_comp}.
%\item[2] Masses in $\log_{10}(M/M_{\sun})$
%\item[3] Within a 4\,kpc radius
\end{tablenotes}
\end{threeparttable}
\label{table:4values}
\end{center}
\end{table*}

\begin{figure*}
\begin{center}
\includegraphics[width=.45\linewidth]{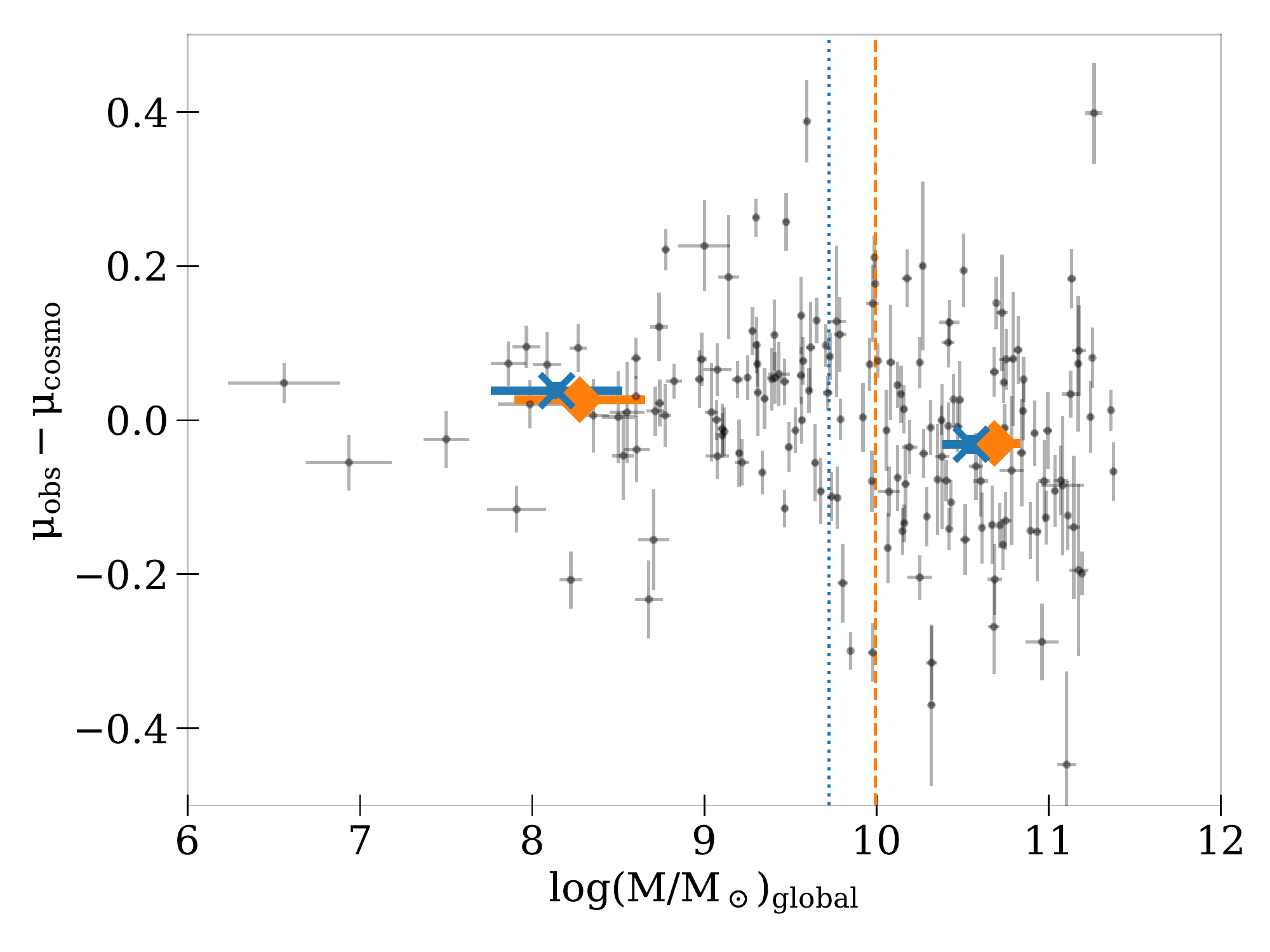}
\includegraphics[width=.45\linewidth]{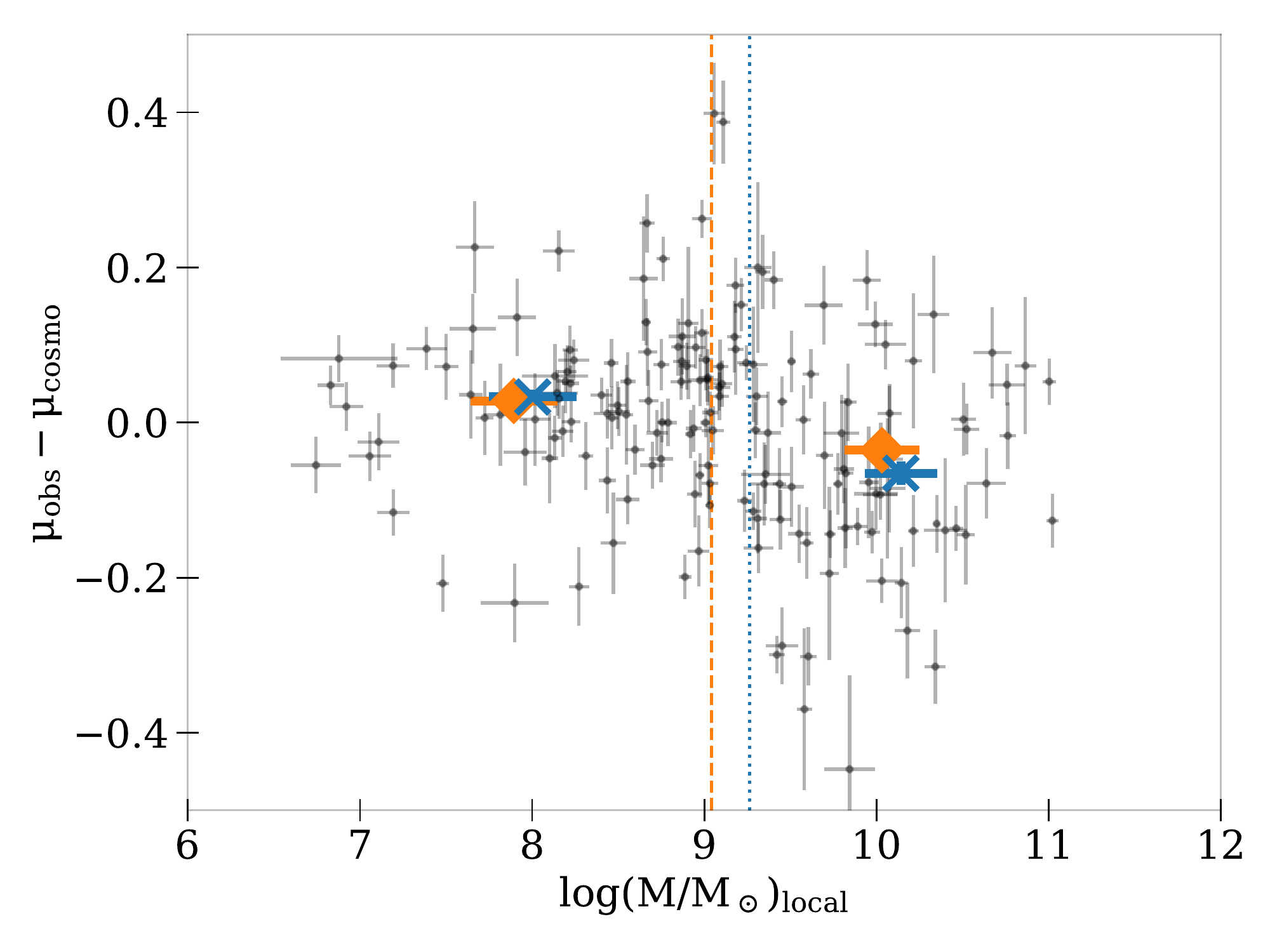}
\includegraphics[width=.45\linewidth]{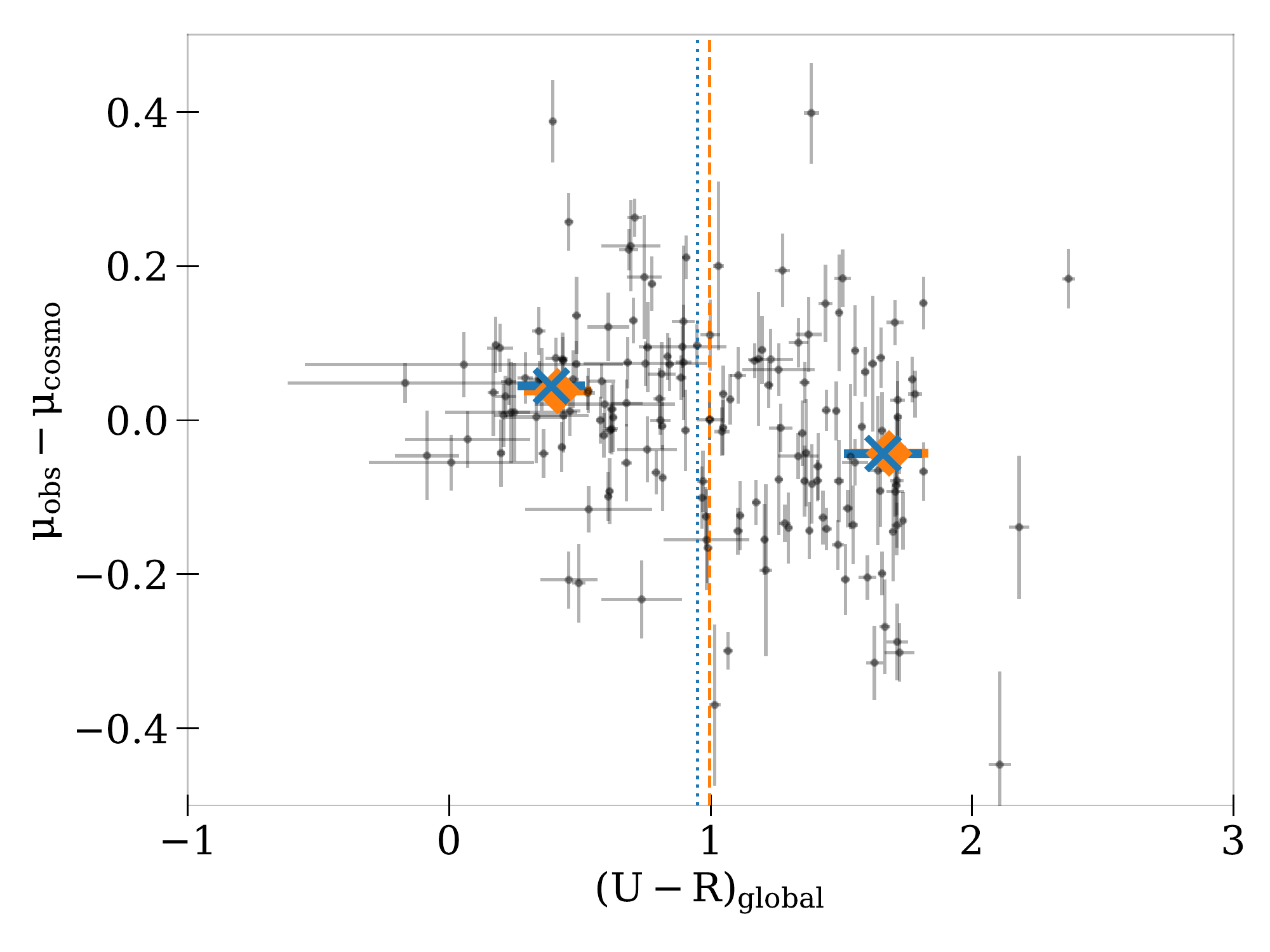}
\includegraphics[width=.45\linewidth]{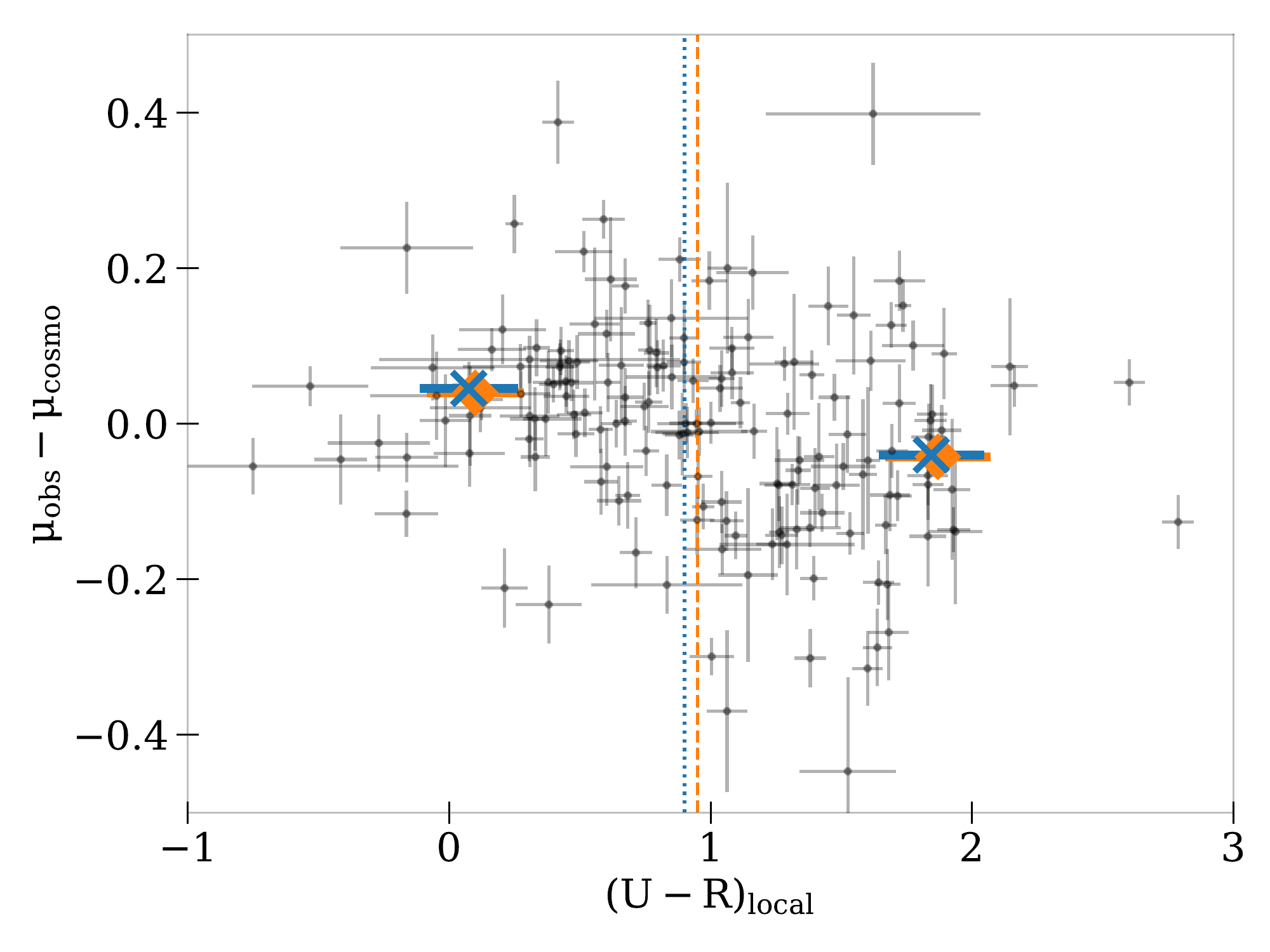}
\caption{Hubble residual plots as a function of (from top to bottom and left to right): global \mstellar, local \mstellar\ within the 4\,kpc radius aperture, global rest-frame $U-R$ colour and local rest-frame $U-R$ colour. The orange-dashed lines represent the sample environmental property medians, and the blue-dotted lines the division point giving the maximum step sizes. These lines correspond with the orange diamond and blue cross bin mean markers, with x-axis error bars showing the dispersion divided by the square root of the number of objects in the bin (as in Figure~\ref{fig:prop_comp}). See Table~\ref{table:4values} for numerical values.}
\label{fig:hubble_mass_colour_step}
\end{center}
\end{figure*}

\begin{figure*}
    \centering
    \includegraphics[width=.45\linewidth]{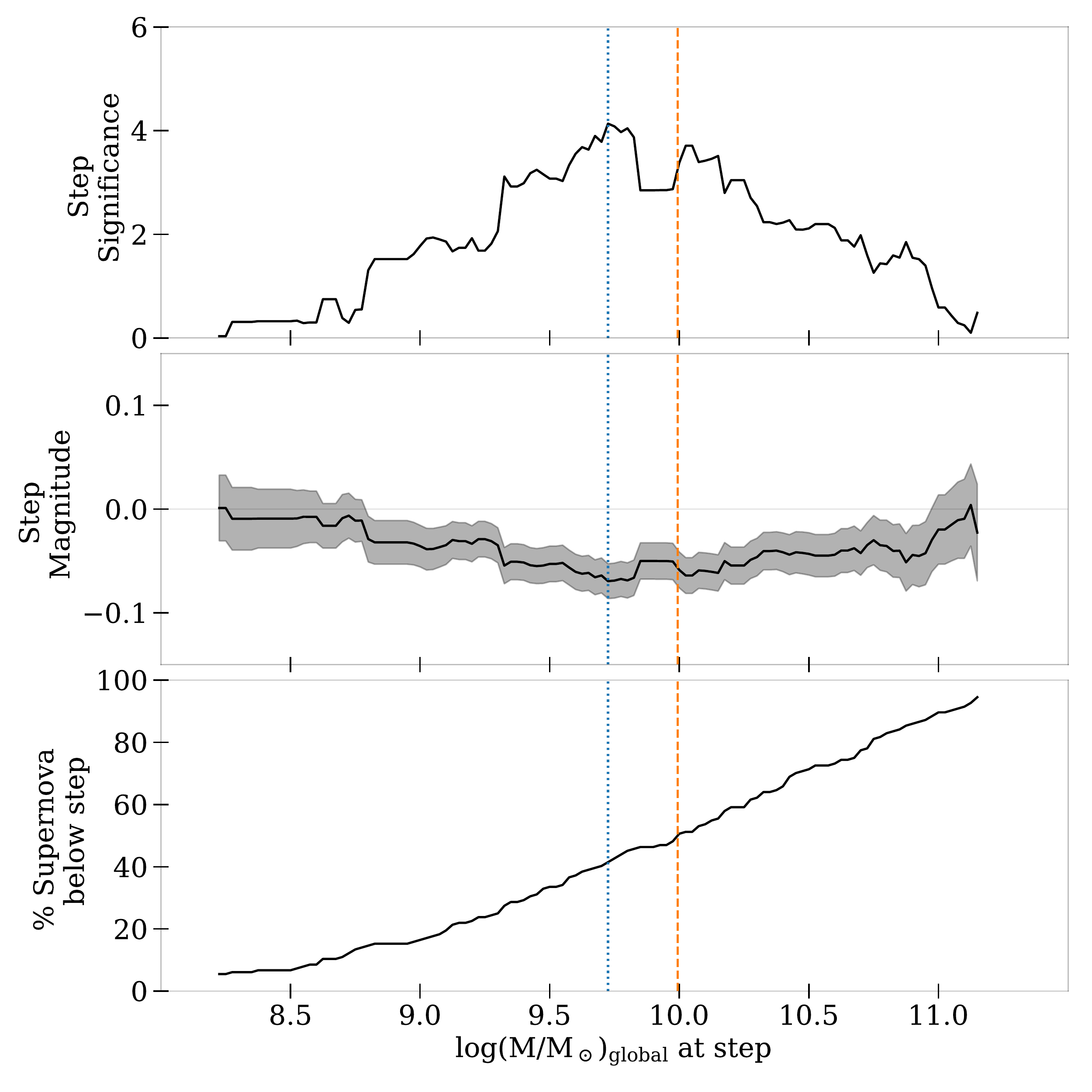}
    \includegraphics[width=.45\linewidth]{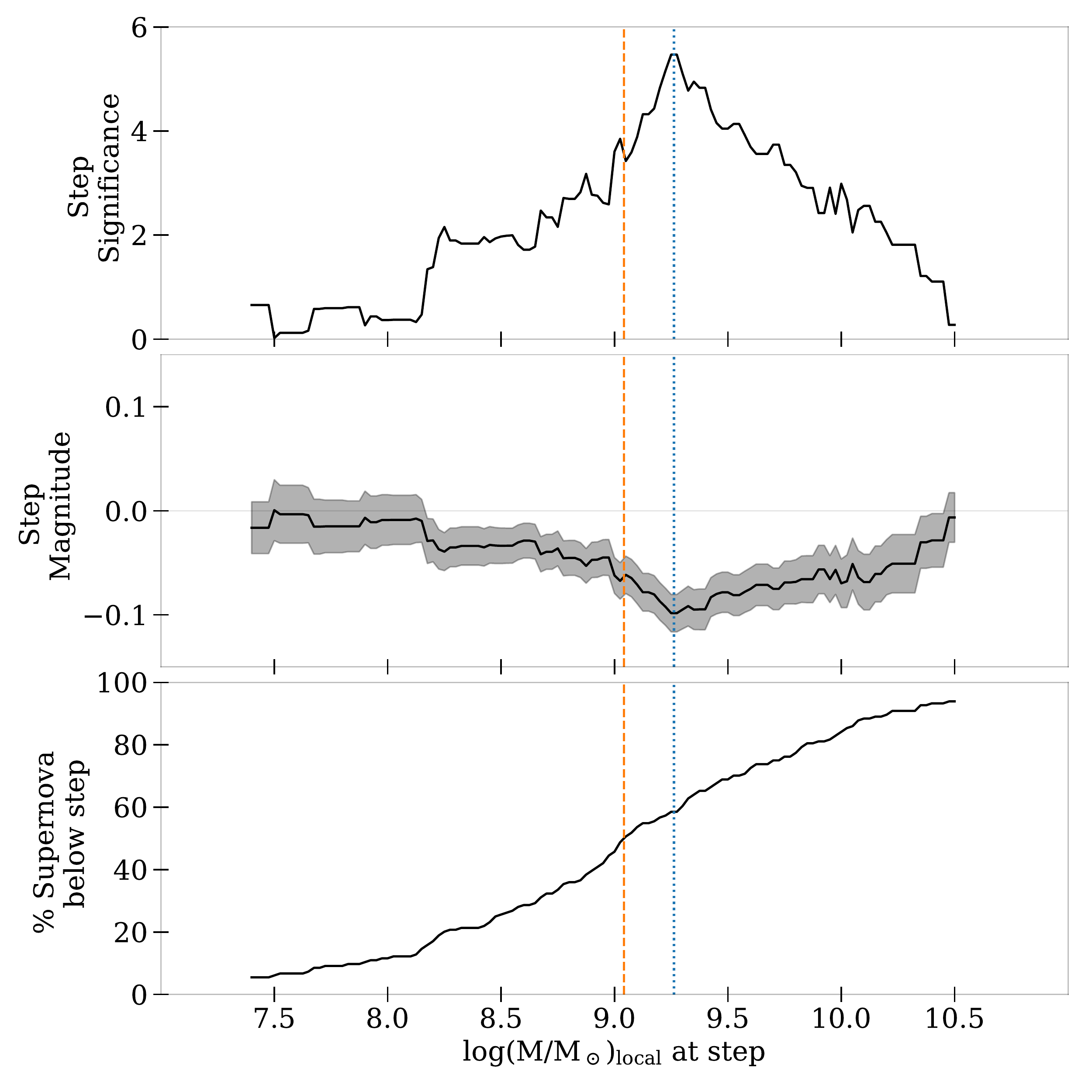}
    \includegraphics[width=.45\linewidth]{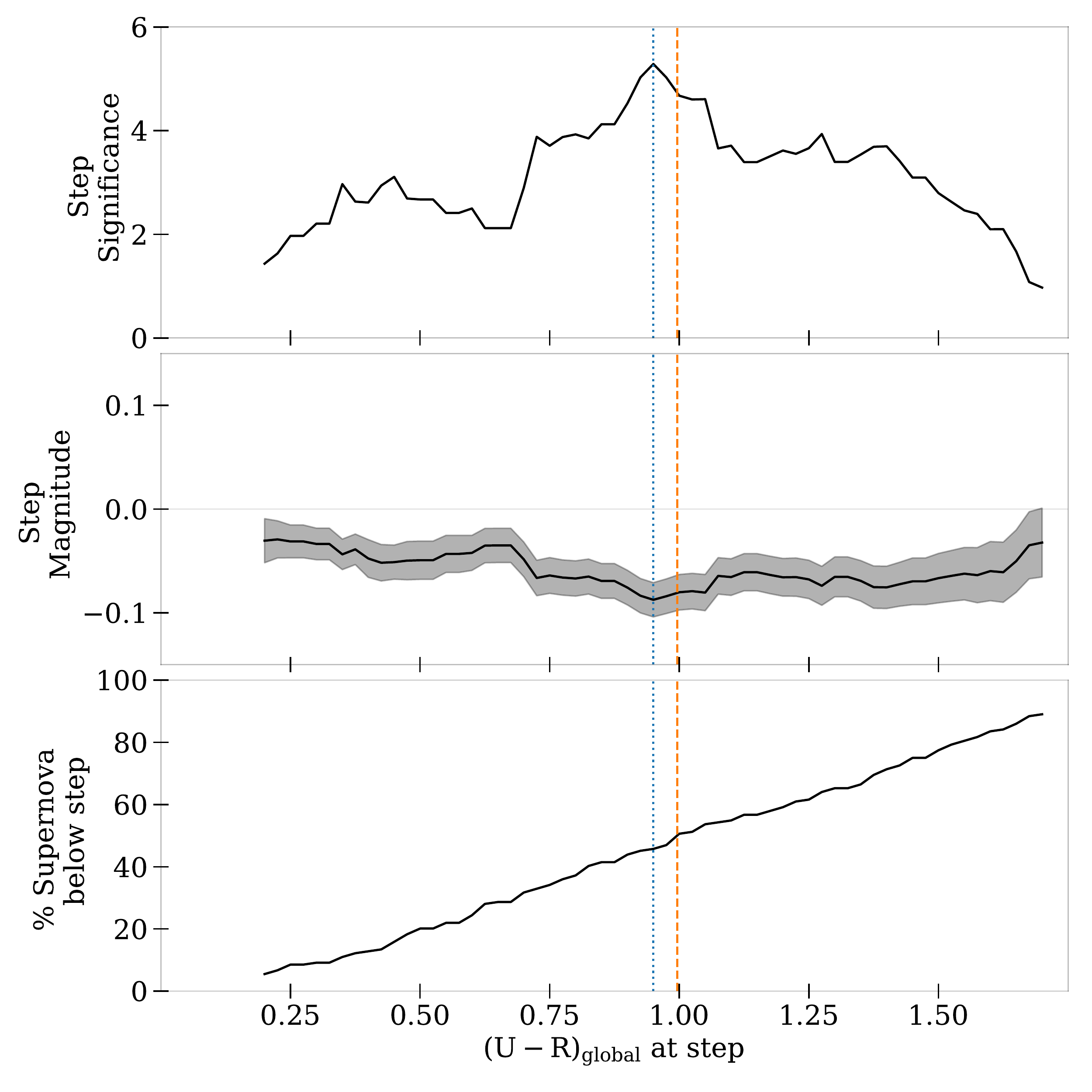}
    \includegraphics[width=.45\linewidth]{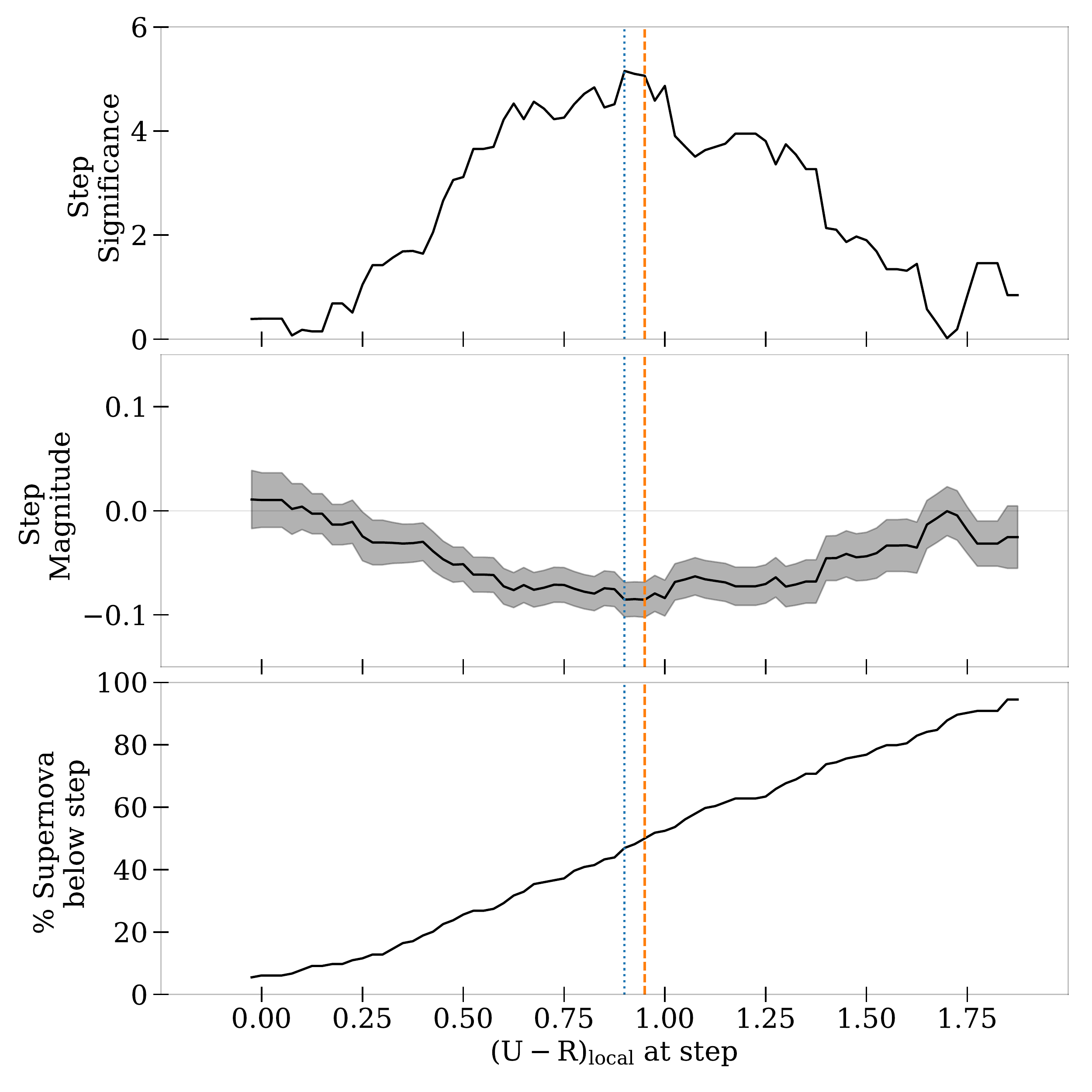}
    \caption{Plots comparing the significance, magnitude and location of the steps for each parameter. From top to bottom, and left to right: global \mstellar, local \mstellar\ within a 4\,kpc radius, global rest-frame $U-R$ colour, and local rest-frame $U-R$ colour. In each plot, the lower panel shows the percentage of SNe Ia in the sample in the bin below the step location as the location of the step is varied; the middle panel is the magnitude of the step at each location with the grey shaded region showing the uncertainty; and the top panel shows the significance of the step in $\sigma$. The orange-dashed line indicates the location of the environmental property median of the sample, and the blue-dotted line shows the step that gives the maximum significance.}
    \label{fig:sig_comp}
\end{figure*}

We next investigate the dependence of SN Ia Hubble residuals on the global and local \mstellar\ and the rest-frame $U-R$ colour of their host galaxies (Fig.~\ref{fig:hubble_mass_colour_step}). As in prior literature, we plot the Hubble residual vs. our chosen host or local property split into two bins at the division point, and measure the mean and dispersion in Hubble residual for environments either side of the division. The magnitude of the step is taken as the difference between these two means. The magnitudes and significances of the step, and r.m.s. values of the Hubble residuals on either side of the step are in Table~\ref{table:4values}. We also explore this SN/host connection across a range of division points in Fig.~\ref{fig:sig_comp}, showing the step locations, significances and magnitudes for the global and local \mstellar\ and $U-R$ steps.

All the measured steps are significant at $>3\sigma$, whether using local or global measures, or using \mstellar\ or $U-R$ colour. The local \mstellar\ step is more significant than the global \mstellar\ step, peaking in significance at a maximum step of $0.098\pm0.018$\,mag. This is around $0.03$\,mag larger than the largest global \mstellar\ step in our sample. We note that the local and global step uncertainties quoted here and in the tables are statistical only. In addition, the complicated positive covariance between the local and global \mstellar\ and local and global $U-R$ colour measures (see Fig.~\ref{fig:z_vs_diff}) will likely increase the significance of the difference in the step size between local and global samples, beyond that obtained with a naive quadratic sum.

We find a local $U-R$ step of $0.082 \pm 0.017$\,mag $(4.8\sigma)$ at the median $U-R$ of the sample, similar in magnitude to that found by \citet{2018A&A...615A..68R} of $0.091 \pm 0.013$\,mag $(7\sigma)$ within a 3\,kpc radius aperture for a larger sample size. At the step with the maximum significance, our $U-R$ step is similar ($0.085\pm0.017$\,mag; 5.2$\sigma$).

The maximum step location and environmental property median step location for the rest-frame $U-R$ colour for both the global and local measurements are located close together, at just below a $U-R$ value of $1.0$, while the \mstellar\ step is more than 1\,dex different. The local $U-R$ measurement has a relatively broad peak (Fig.~\ref{fig:sig_comp}; i.e., the step size is insensitive to the split point). This may suggest that the local $U-R$ step is more stable -- but perhaps also less discriminating -- than the local \mstellar\ step. This is consistent with \citet{2018A&A...615A..68R}, who found similar magnitude steps for global and local rest-frame $U-V$ colour (their table 7). 

From Table~\ref{table:4values} we also note that the r.m.s. values for the Hubble residuals are smaller in bluer galaxies/environments and lower mass galaxies/environments (cf. Section~\ref{prop_v_host}) by an average of 1.3$\sigma$. 

\begin{figure*}
    \centering
    \includegraphics[width=\linewidth]{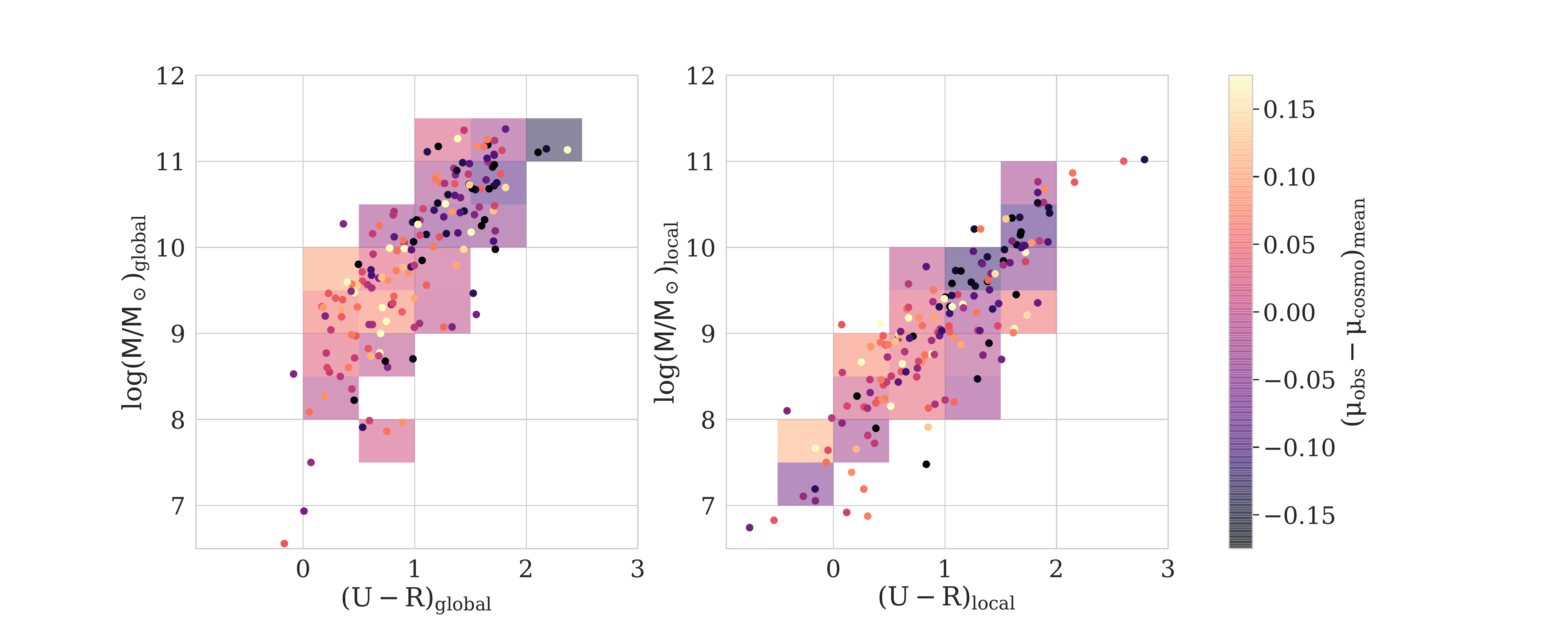}
    \caption{Heatmaps displaying correlations between rest-frame $U-R$ and stellar mass, with bins weighted by mean Hubble residual, for both global (left plot) and local (right plot) environments. Bins containing two or fewer SN are not displayed. Overplotted with scatter plot displaying raw, unbinned data. To give an indication of the uncertainty in the colour coding, we quote the median uncertainty in the bin mean Hubble residuals as 0.037\,mag.}
    \label{fig:resi_heatmap}
\end{figure*}

In Fig.~\ref{fig:resi_heatmap}, we show a complementary visualisation of our data using two-dimensional heatmaps in the parameter space of rest-frame $U-R$ and stellar mass, with bins in this space coloured by mean Hubble residual. This visualisation allows an examination of trends in Hubble residual with a given host galaxy property, at a fixed value of a different host galaxy property; for example, the variation in Hubble residual with $U-R$ colour at fixed stellar mass.

Any variation in Hubble residual at fixed environmental property is quite minimal, but as an example,  keeping global stellar mass constant just below $10^{10}$\,M$_{\sun}$, we see a very slight decrease in Hubble residual with increasing global $U-R$. With a larger sample, we will investigate this further and the effect of combining environmental properties. 

\section{Systematics} \label{sys}

In this section, we perform additional tests to explore some of our analysis choices and their effect on the results. We study the effect of changing the size of the local aperture photometry radius, the cosmological bias correction, and the use of sSFR in place of $U-R$. 

\subsection{Changing local radius} \label{varying}

Our main analysis uses a local physical radius of 4\,kpc. Here, we vary the size of this local radius, choosing aperture radii of size from 2.5 to 10\,kpc in 0.5\,kpc steps, and from 10 to 30\,kpc in 5\,kpc steps. This probes a wide range from very small apertures to those of galactic size.

For all apertures, we follow the original method, remeasuring all derived galaxy parameters. As a result, the number of objects may vary slightly for objects near the boundaries of the cuts discussed in Section~\ref{selection_cuts}. For example, at the smallest radii there are fewer photons entering the aperture leading to larger statistical uncertainties in the measurements of properties, and thus more objects are likely to be rejected. We also note that in the largest radii, satellite, companion, or background galaxies may additionally enter the aperture. This effect is redshift dependent: a 30\,kpc aperture at $z=0.1$ is likely to have more background sources than one at $z=0.5$.

\begin{figure*}
\begin{center}
\includegraphics[width=.45\linewidth]{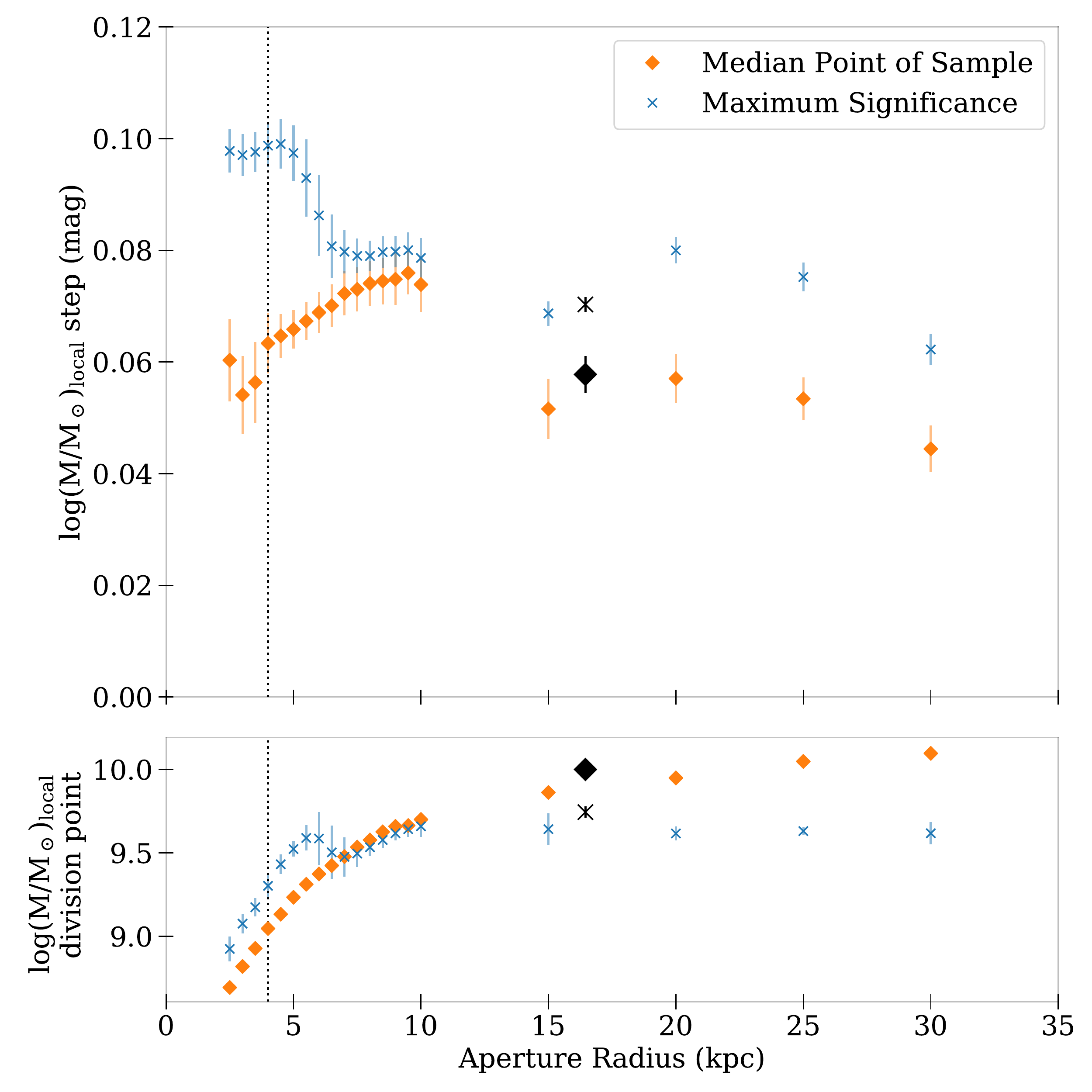}
\includegraphics[width=.45\linewidth]{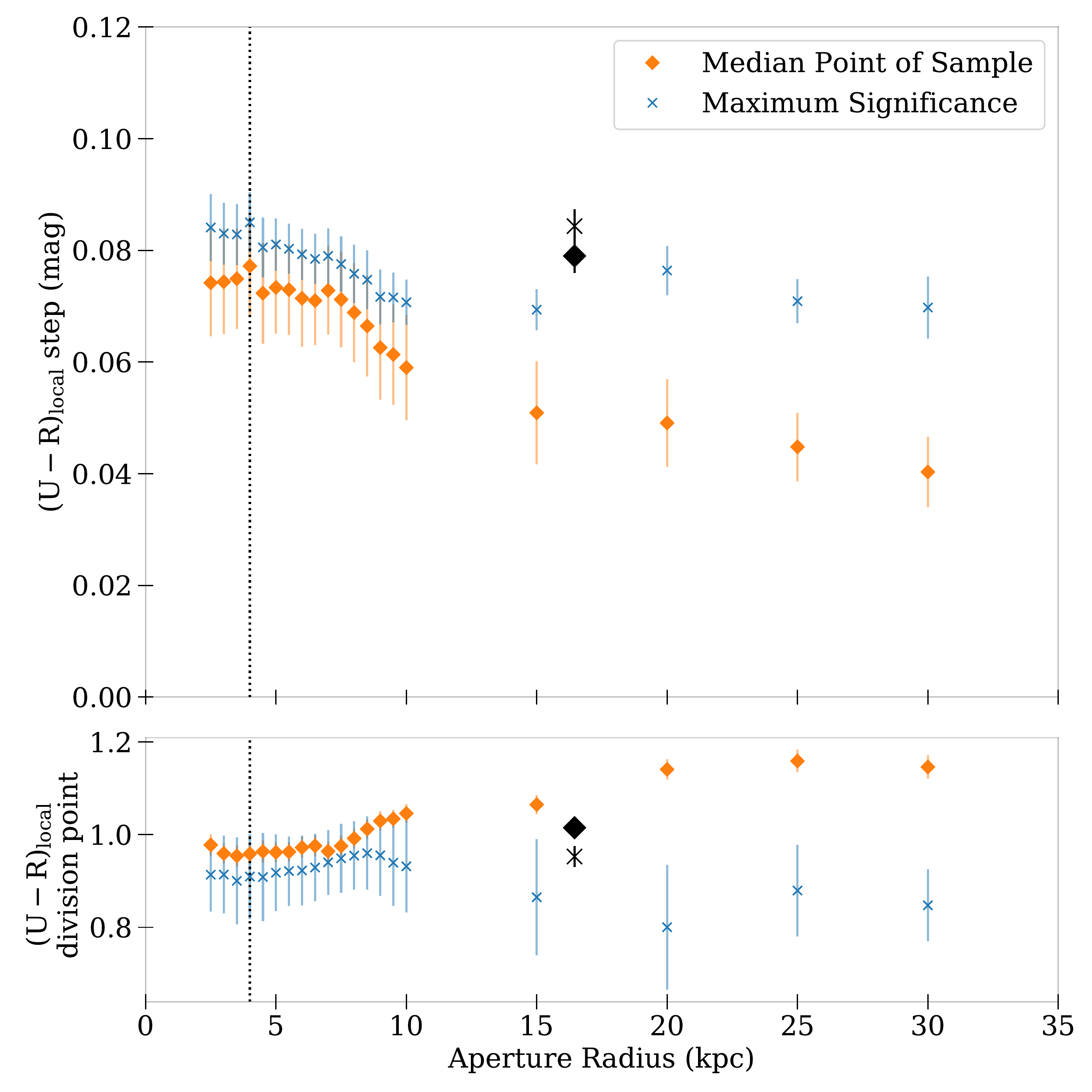}
\caption{Top panels: variation of the local \mstellar\ step and the rest-frame $U-R$ colour step as functions of the local aperture radius. Lower panels: evolution of the division points of the step as a function of the local aperture radius. Error bars in all panels are the standard deviations of the sample, however we acknowledge that there will be correlations between the different aperture sizes. The global measurements for both \mstellar\ and $U-R$ are represented by the black symbols (diamond for median, cross for point of maximum significance) placed at the average sized aperture radius for the sample used of $16.45\pm6.08$\,kpc. The black dotted line in all panels indicates the fiducial 4\,kpc aperture sized used in our analysis. This additionally is the minimum size that includes all data points, results below this aperture are potentially biased by the PSF size.}
\label{fig:step_u_r} 
\end{center}
\end{figure*}

Fig.~\ref{fig:step_u_r} shows the local $U-R$ and local \mstellar\ step as a function of aperture radius. The magnitude of the step decreases as the aperture size increases. We note, particularly for the $U-R$, that the global measurements are not following this trend; the global result is not the asymptotic limit. This is likely to be due to the difference in aperture types used: for the iteration over different aperture sizes, circular apertures were used, whereas the global measurements use a Kron-like aperture. In addition, as global properties of galaxies are centred on the galaxy centre and local measurements are centred on the individual SN locations, these measurements may never converge.   

In order to calculate the average global radius of the host galaxies in our sample we used the \textsc{Source Extractor} output values to obtain an area for each measured ellipse, equating this to the area of a circle, thus obtaining an effective circular aperture radius in arcseconds for each host galaxy, which can be compared to the circular aperture radii used in our local analysis rather than the semi-major and semi-minor axes of an ellipse. We convert these angular distances to proper distances for each host, and find that the average circular galaxy radius in our sample is $16.45\pm6.08$\,kpc. Hence, for the large local-aperture radii where the shape of the circular aperture extends over the edge of the galaxy measured by \textsc{Source Extractor}, there may be a greater effect on the $U-R$ measurement as the apertures includes background flux not associated with the host. This may introduce additional scatter in the overall colour measurement of the region, and thus affect the Hubble residual dependence.

The local $U-R$ measurement is more consistent than the local \mstellar, exhibiting less of a difference between the step magnitudes when making the division point at the environmental property median or maximum step significance point of the sample. Furthermore, from the lower panels of Fig.~\ref{fig:step_u_r}, whilst as expected the local mass displays a decreasing division point as smaller apertures are used (and less mass is contained in the aperture), the local $U-R$ division point for both the environmental property median point of the sample and the location of the maximum significant step is  consistent at a value of $\sim1$ for all apertures below 15\,kpc in radius. This makes a cosmological analysis using $U-R$ as a probe of environment characteristics simpler, as it is less dependent on the local aperture radius size and the location of the division point, and the step location can  be set at $U-R = 1$. This finding suggests that the local $U-R$ is more stable than \mstellar. 

\subsection{5D or 1D Cosmological Corrections} \label{vary_cosmo}

As discussed in Section~\ref{params}, we use the 1D bias correction in our baseline analysis, as opposed to the 5D $\mu_\mathrm{bias}$ BBC correction.

If we use the 5D $\mu_\mathrm{bias}$ correction, we find that the magnitudes of Hubble residual steps are, as expected, smaller than those for the 1D corrections by an average of $\sim0.026$\,mag across \mstellar\ and $U-R$, as shown in Table~\ref{table:1dv5d}. This difference is likely due to the effects of the underlying simulated $x_1$-\mstellar\ correlation that is not modelled in existing 5D corrections, with \citetalias{2020MNRAS.494.4426S} finding a similar difference of $0.026\pm0.009$\,mag in their analysis. 

\begin{table*}
\begin{center}
\caption{As Table~\ref{table:4values}, but for a 5D bias correction.}

\begin{tabular}{c c c c c }

\hline 
\multicolumn{1}{c}{Property} & \multicolumn{1}{c}{Sample Median/} & \multicolumn{1}{c}{Division} & \multicolumn{2}{c}{Hubble Residual} \\&  \multicolumn{1}{c}{Max Significance} & \multicolumn{1}{c}{Point} & Sig. ($\sigma$) & Magnitude\\

\hline

Global Mass & Median & 9.99 & 1.06 & 0.019 $\pm$ 0.018  \\ 
Global Mass & Max & 9.73 & 2.17 & 0.037 $\pm$ 0.017 \\ 
Local Mass & Median & 9.04 & 2.36 & 0.042 $\pm$ 0.018  \\ 
Local Mass & Max & 9.26 & 4.34 & 0.076 $\pm$ 0.017  \\ 
\hline
Global U-R & Median & 1.00 & 3.37 & 0.058 $\pm$ 0.017  \\ 
Global U-R & Max & 0.95 & 3.77 & 0.064 $\pm$ 0.017 \\ 
Local U-R & Median & 0.95 & 3.16 & 0.055 $\pm$ 0.017  \\ 
Local U-R & Max & 0.95 & 3.82 & 0.065 $\pm$ 0.017  \\ 
\hline

\hline
\end{tabular}
\label{table:1dv5d}
\end{center}
\end{table*}

To address this 1D vs. 5D difference, future DES work will include $x_1$-\mstellar\ correlations in the simulations, and a BBC dependence on \mstellar.

\subsection{Use of sSFR}

We have focused on \mstellar\ and rest-frame $U-R$ colour, but additional galaxy properties are available from the SED fitting code described in Section~\ref{SED}, in particular the sSFR. We find that the local sSFR step is of similar magnitude and significance to the local $U-R$ step measurements, with the local sSFR step found at the sample median sSFR location being $0.064 \pm 0.017$\,mag $(3.7\sigma)$ for the 4\,kpc radius aperture between star-forming and passive regions. This is consistent with the $0.081 \pm 0.018$\,mag step found by \cite{2019JKAS...52..181K}, but considerably smaller than the $0.163 \pm 0.029$\,mag step found by \cite{2018arXiv180603849R}, although we note that our measure of sSFR is less direct (based on template fitting), and \cite{2018arXiv180603849R} is at low $z$, where we expect the step to be larger; see \cite{2013A&A...560A..66R} Figure 11, also \cite{2014MNRAS.445.1898C} and \cite{2018ApJ...854...24K}.

\section{Discussion} \label{discussion}

\subsection{The use of different rest-frame colours} \label{colours}

Using the rest-frame $UBVR$ magnitudes calculated in Section~\ref{SED} for each global host galaxy and local region, we can measure a variety of rest-frame colours. We chose $U-R$ for our main analysis as it covers the largest wavelength range for our DES dataset, traces both the red and blue ends of the spectrum so carrying information about both \mstellar\ and SFR (and thus age) of the stellar populations, and has been found to correlate with galaxy morphology \citep{2008MNRAS.389.1179L}. For completeness, we present all rest-frame colour steps across all radii studied in the Online Supplementary Material. 

We note that the largest local rest-frame colour step of $0.099 \pm 0.016$\,mag ($6\sigma$) for the 4\,kpc radius aperture is found when using $V-R$ colour. $V-R$ represents two neighbouring filter responses in our $UBVR$ estimates. Further investigation is needed with a larger dataset to determine which colour is the most stable and effective for use in cosmological analysis. 

\subsection{Splitting the sample by stretch and colour}

Although the DES3YR sample is of modest size, we perform a preliminary investigation of splitting the sample by SN $x_1$ and $c$: $x_1>0$ and $x_1\leq 0$, and $c > 0$ and $c\leq0$, following \cite{2010MNRAS.406..782S} and \cite{2018arXiv180603849R}. This tests whether the steps in SN Ia luminosity could be driven by underlying relationships between $x_1$/$c$ and host galaxy properties.

We repeat our analysis using these subsamples, and present in Table~\ref{table:x1_c_split} the step magnitudes and uncertainties for steps at the median environmental property ($U-R$ and \mstellar) division point of the sample. We use our default 4\,kpc radius aperture and a 1D cosmological bias correction throughout. 

\begin{table*}
\begin{center}
\caption{Subsample data when splitting the sample based on on $x_1$ and $c$.}
\begin{threeparttable}
\begin{tabular}{c c c c c c c} 
\hline
\multicolumn{1}{c}{Property} & \multicolumn{3}{c}{$c$ split} & \multicolumn{3}{c}{$x_1$ split} \\&  $c < 0$ & $c > 0$ & Difference ($\sigma$)\tnote{1} & $x_1 < 0$ & $x_1 > 0$ & Difference ($\sigma$)\\
\hline
Number of Supernovae & 102 & 62 &  & 74 & 90 & \\
\hline
\multicolumn{7}{l}{Keeping $\alpha$ and $\beta$ fixed ($\alpha = 0.156 \pm 0.012$, $\beta = 3.201 \pm 0.131$)} \\
Global Mass Step\tnote{2} & $ 0.012 \pm 0.020 $ & $ 0.141 \pm 0.029 $ & 3.66 & $ 0.006 \pm 0.026 $ & $ 0.064 \pm 0.024 $ & 1.64 \\
Local Mass Step & $ 0.034 \pm 0.020 $ & $ 0.125 \pm 0.031 $ & 2.47 & $ 0.069 \pm 0.025 $ & $ 0.041 \pm 0.025 $ & 0.79 \\
Global U-R Step & $ 0.045 \pm 0.020 $ & $ 0.154 \pm  0.029 $ & 3.09 & $ 0.065 \pm 0.024 $ & $ 0.070 \pm 0.024 $ & 0.14 \\
Local U-R Step & $ 0.046 \pm 0.020 $ & $ 0.148 \pm 0.029 $ & 2.89 & $ 0.058 \pm 0.025 $ & $ 0.085 \pm 0.023 $ & 0.79\\

%\hline
%\multicolumn{7}{l}{$\alpha$ and $\beta$ free to best-fit for each sub-sample} \\
%$\alpha$ & $ 0.158 \pm 0.017 $ & $ 0.094 \pm 0.028 $ & 1.95 & $ 0.254 \pm 0.028 $ & $ 0.228 \pm 0.031 $ & 0.62 \\
%$\beta$ & $ 4.487 \pm 0.619 $ & $ 3.894 \pm 0.481 $ & 0.76 & $ 3.273 \pm 0.195 $ & $ 3.585 \pm 0.211$ & 1.09 \\
%Global Mass Step & $ 0.033 \pm 0.026 $ & $ 0.123 \pm 0.033 $ & 2.14 & $ 0.030 \pm 0.027 $ & $ 0.069 \pm 0.027 $ & 1.02 \\
%Local Mass Step & $ 0.049 \pm 0.026 $ & $ 0.126 \pm 0.034 $ & 1.79 & $ 0.088 \pm 0.025 $ & $ 0.066 \pm 0.027 $ & 0.60 \\
%Global U-R Step & $ 0.071 \pm 0.026 $ & $ 0.143 \pm 0.032 $ & 1.75 & $ 0.094 \pm 0.025 $ & $ 0.079 \pm 0.026 $ & 0.42\\
%Local U-R Step & $ 0.077 \pm 0.025 $ & $ 0.118 \pm 0.034 $ & 0.97 & $ 0.097 \pm 0.025 $ & $ 0.103 \pm 0.025 $ & 0.17\\
\hline 
\end{tabular}
\begin{tablenotes}
\item[1] Significance is quadrature sum.
\item[2] Step division point at the subsample environmental property ($U-R$ and \mstellar) median.
\end{tablenotes}
\end{threeparttable}
\label{table:x1_c_split}
\end{center}
\end{table*}

When $\alpha$ and $\beta$ are fixed at the values derived from the full DES3YR sample ($\alpha = 0.156 \pm 0.012$, $\beta = 3.201 \pm 0.131$), we find a significant difference ($\sim 3\sigma$) between step sizes for subsamples split for high and low $c$, as displayed in Table~\ref{table:x1_c_split}. We find that the bluer $c < 0$ have smaller steps than for $c > 0$, indicating that the bluer subset is more homogeneous. The redder $c > 0$ have higher dispersion and larger steps ($\sim0.14$\,mag), similar to in \citetalias{2020MNRAS.494.4426S}. This is not consistent with \citet{2018arXiv180603849R} who found no significant difference between the size of the Local sSFR bias in subsamples split for high and low $c$ ($0.45\sigma$). We find no significant difference between the step sizes for the subsamples split into high and low $x_1$, consistent with \citet{2010MNRAS.406..782S}, who found no significant difference between the size of the global stellar mass step in subsamples split for high and low $x_1$, with an average difference between subsamples of $0.60\sigma$; and with \citet{2018arXiv180603849R} who found a difference of $0.84\sigma$.  

There are two broad interpretations of our findings for the subsamples split for $c$: redder and bluer objects may represent different progenitor paths, i.e. bluer objects represent one distinct set of progenitors (hence no step), whilst redder objects are a combination of different progenitors (hence show a step); or the bluer $c < 0$ objects may suffer less dust extinction \citep{2020arXiv200410206B} and thus less event-to-event scatter. These interpretations suggest that the Hubble residual steps that we see in the main sample may be driven by physics that effects the colour of the SNe, and could be a source for the origin of the remaining $\sim0.14$\,mag \citep{2018ApJ...859..101S} Hubble residual dispersion in the general SN Ia population. Alternatively, as bluer objects are observationally brighter than redder objects and so have lower uncertainties, they drive the fit of $\alpha$ and $\beta$ so they may in turn drive the size of the step.

\begin{figure*}
\begin{center}
    
\includegraphics[width=8.25cm]{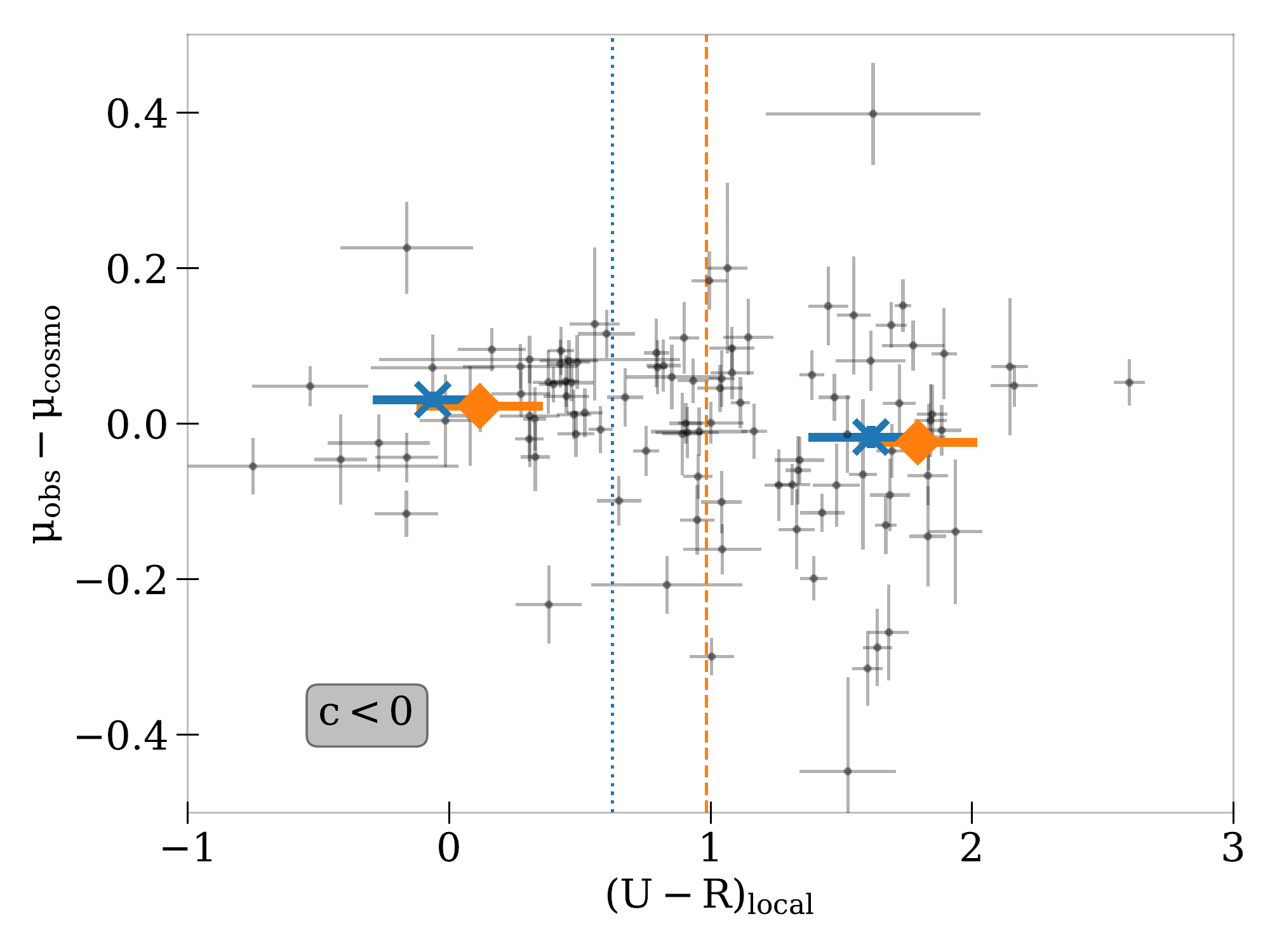}
\includegraphics[width=8.25cm]{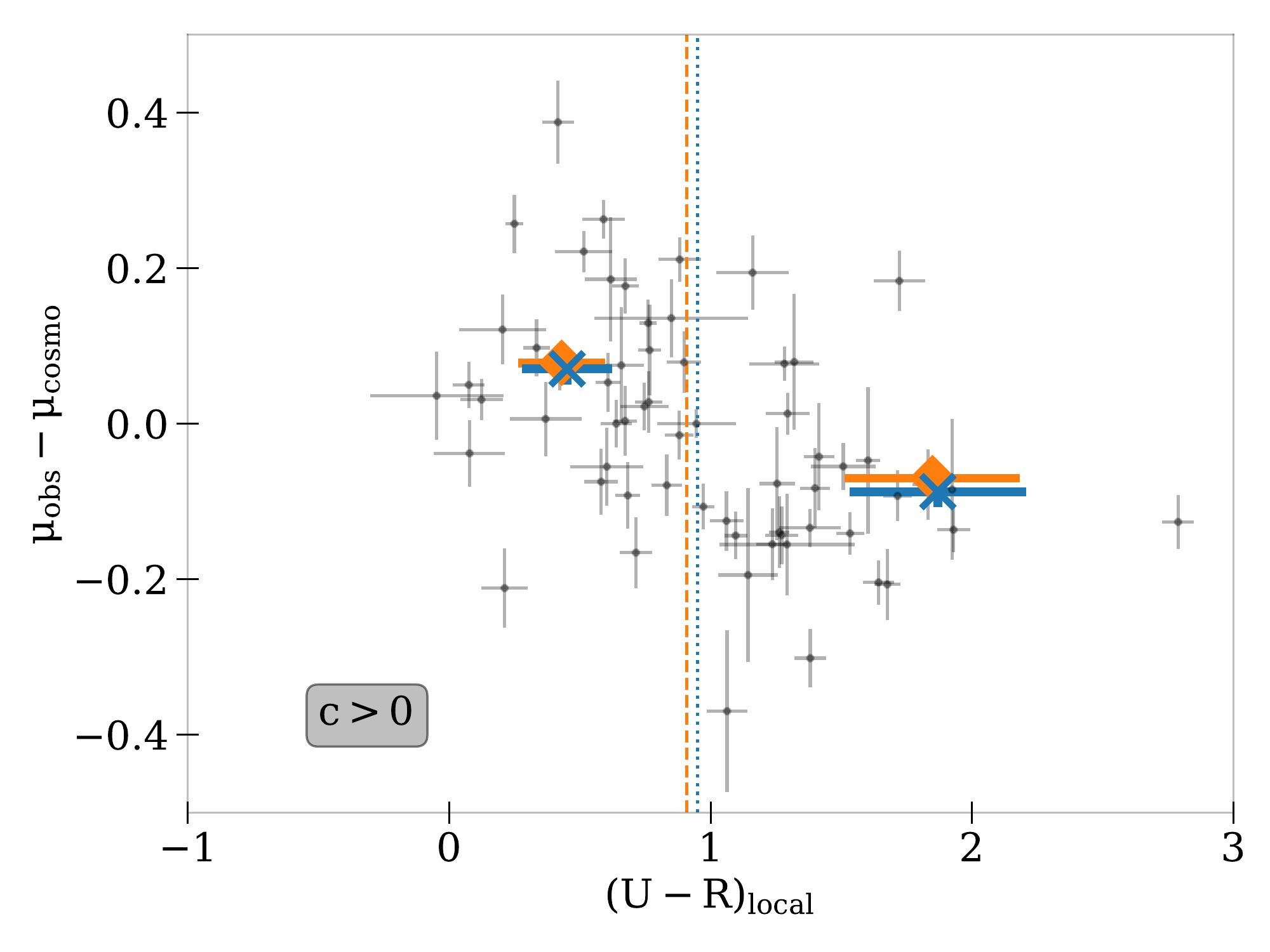}
\includegraphics[width=8.25cm]{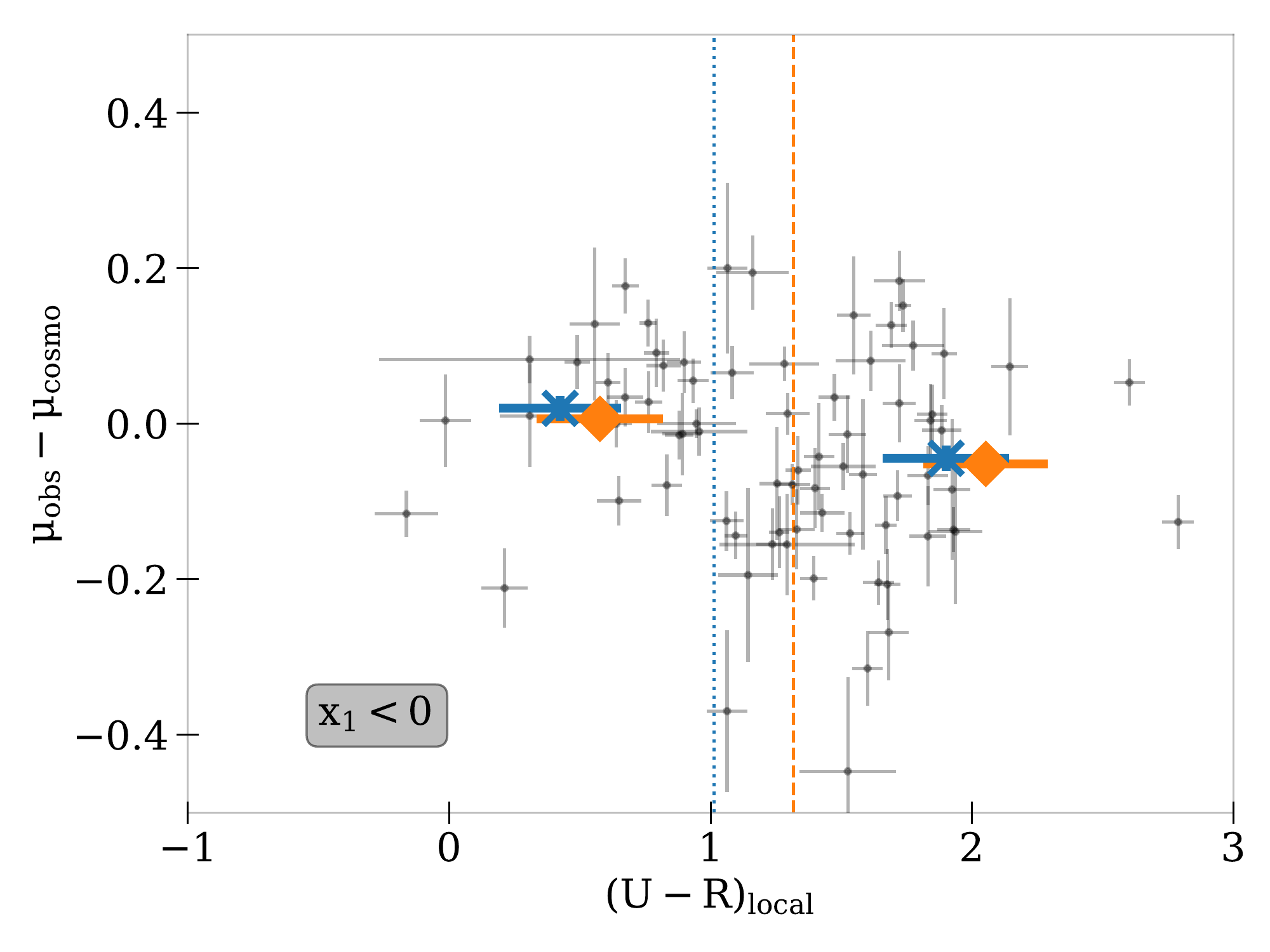}
\includegraphics[width=8.25cm]{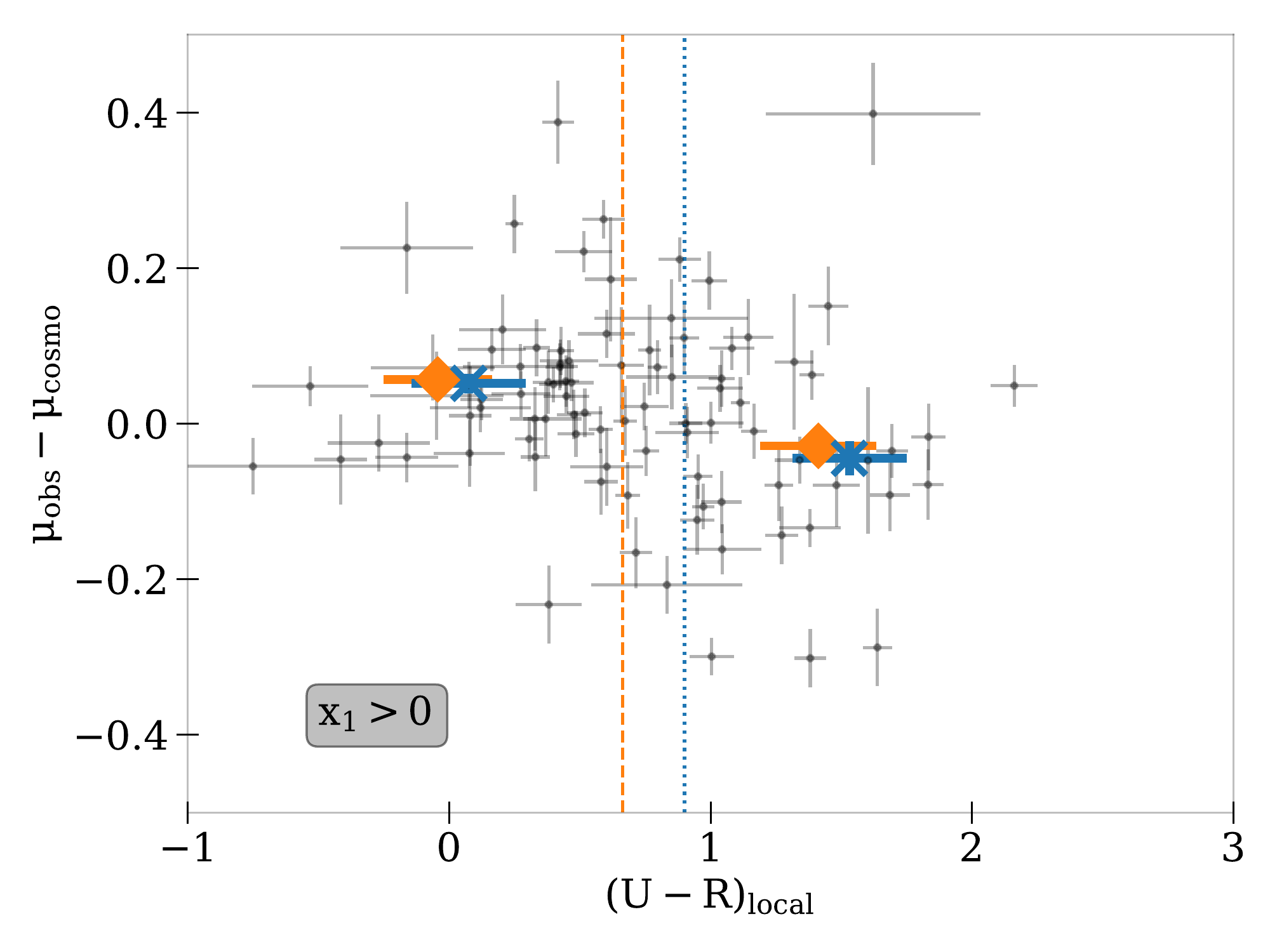}

\caption{Hubble residual plots as a function of local $U-R$ within the 4\,kpc radius aperture, for subsamples split by $c$ and $x_1$, where $\alpha$ and $\beta$ have been fixed as in the main analysis. As in Fig~\ref{fig:hubble_mass_colour_step}, the orange-dashed line represent the sample environmental property median, and the blue-dotted line the location of the maximum step. These correspond with the orange diamond and blue cross bin mean markers. See Table~\ref{table:x1_c_split} for the numerical values for the steps when split at the environmental property median of the subsamples. r.m.s. values are displayed in Table~\ref{table:rms_U-R_splitsample}.}
\label{fig:x1_c_split_hubble_comp}

\end{center}
\end{figure*}

%RMS values for U-R local (fig 11)

%not sure I need both median and max values - make table too busy?

%c_l_0-origab
%RMS median left: 0.08445839984638709
%RMS median right: 0.14808896212472794
%RMS max left: 0.08098963229606988
%RMS max right: 0.13673308251287525
%c_g_0-origab
%RMS median left: 0.14136067878448325
%RMS median right: 0.15041082802028644
%RMS max left: 0.13885717765907532
%RMS max right: 0.15357687437764816
%x1_l_0-origab
%RMS median left: 0.12331023605454638
%RMS median right: 0.14730361290851382
%RMS max left: 0.08852045404420966
%RMS max right: 0.15238970730813026
%x1_g_0-origab
%RMS median left: 0.11573391811898802
%RMS median right: 0.13626914257593878
%RMS max left: 0.11691023291918103
%RMS max right: 0.1412951171625869

\begin{table}
\begin{center}

\caption{RMS values for local U-R for the split subsamples, corresponding to Fig.~\ref{fig:x1_c_split_hubble_comp}, and Table.~\ref{table:x1_c_split}.}
\begin{threeparttable}
\renewcommand{\TPTminimum}{\linewidth}
\makebox[\linewidth]{%
\begin{tabular}{c c c }

\hline 
Sub-sample\tnote{1}& \multicolumn{2}{c}{Local U-R RMS} \\
&  < DP\tnote{2} & > DP \\
\hline
$c < 0$ & 0.084 $\pm$ 0.017 & 0.148 $\pm$ 0.030\\
$c > 0$ & 0.141 $\pm$ 0.037 & 0.150 $\pm$ 0.039\\
$x_1 < 0$ & 0.123 $\pm$ 0.029 & 0.147 $\pm$ 0.035\\
$x_1 > 0$ & 0.116 $\pm$ 0.025 & 0.136 $\pm$ 0.029\\
\hline
\end{tabular}}
\begin{tablenotes}
\item[1] Subsamples where $\alpha$ and $\beta$ were kept fixed ($\alpha = 0.156 \pm 0.012$, $\beta = 3.201 \pm 0.131$). 
\item[2] Division point at subsample local $U-R$ median.
\end{tablenotes}
\end{threeparttable}
\label{table:rms_U-R_splitsample}
\end{center}
\end{table}

The Hubble residual r.m.s. values for local $U-R$ in Table~\ref{table:rms_U-R_splitsample} are similar to our main analysis, with SNe Ia in the bluer galaxies being more homogeneous. This homogeneity is most pronounced when the sample is split by $c$, with a particularly low r.m.s. of $0.084 \pm 0.017$\,mag for blue SNe Ia in blue local galaxy regions. 

When we refit $\alpha$ and $\beta$ to best fit for our $x_1$ and $c$ subsamples our findings are consistent but the sizes of the steps decrease slightly, indicating that the effects of the steps have been absorbed by the changing $\alpha$ and $\beta$ parameters. However, by modifying $\alpha$ and $\beta$ in this way, we are no longer recovering the underlying effect of the $x_1$ and $c$ parameters on the main sample, so this may not be the best representative of the fundamental cause of the steps.

\subsection{Splitting the sample by environmental properties} \label{split_env_refit}

As a final test, we refit $\alpha$ and $\beta$ for subsamples based on splitting by the environmental properties of stellar mass and rest-frame $U-R$. Similarly to splitting by $c$ and $x_1$, this test investigates whether the steps in SN Ia luminosity could be driven by underlying relationships between $x_1$/$c$ and host galaxy properties. 

There is a small, but interesting, $\sim 2\sigma$ difference in both $\alpha$ and $\beta$ values on each side of the environmental property division point (Table~\ref{table:env_split}). In \citet{2020arXiv200410206B}, SNe found in high mass galaxies are suggested to follow a different colour law compared to those in lower mass systems, thus it is expected that they would have a lower $\beta$. We see evidence of this expectation with smaller $\beta$ values found in higher mass, redder regions; agreeing with \citet{2011ApJ...737..102S}. We also see a smaller $\alpha$ for low mass, bluer regions, suggestive of the relationship between host galaxy stellar mass and $x_1$, and the prediction of \citet{2014MNRAS.445.1898C}: that the most cosmologically uniform sample is located in actively star-forming, lower-mass galaxies. The lower $\alpha$ value means that there is less need for a correction for these SNe, therefore they have lower scatter and thus they are better standard candles. As stated, the difference that we find using the three year spectroscopically confirmed sample is small, $\sim 2\sigma$, but motivates further analysis of this tentative result using DES-5YR.  

\begin{table*}
\begin{center}
\caption{Subsample data when splitting the sample based on on environmental properties.}
\begin{threeparttable}
\begin{tabular}{c c c c c c c} 
\hline
\multicolumn{1}{c}{Property} & \multicolumn{3}{c}{$\alpha$} & \multicolumn{3}{c}{$\beta$} \\&  < DP\tnote{2} & > DP & Difference ($\sigma$)\tnote{1} & < DP & > DP & Difference ($\sigma$)\\
\hline
Global Mass & $ 0.140 \pm 0.020 $ & $ 0.198 \pm 0.015 $ & 2.32 & $ 3.65 \pm 0.22 $ & $ 3.16 \pm 0.19 $ & 1.69 \\
Local Mass & $ 0.156 \pm 0.017 $ & $ 0.206 \pm 0.020 $ & 1.90 & $ 3.65 \pm 0.19 $ & $ 3.07 \pm 0.20 $ & 2.10 \\
Global U-R & $ 0.154 \pm 0.021 $ & $ 0.219 \pm  0.018 $ & 2.35 & $ 3.71 \pm 0.18 $ & $ 3.08 \pm 0.20 $ & 2.34 \\
Local U-R & $ 0.157 \pm 0.023 $ & $ 0.218 \pm 0.017 $ & 2.13 & $ 3.62 \pm 0.22 $ & $ 2.99 \pm 0.18 $ & 2.21\\
\hline 
\end{tabular}
\begin{tablenotes}
\item[1] Significance is quadrature sum.
\item[2] Division point at the sample median.
\end{tablenotes}
\end{threeparttable}
\label{table:env_split}
\end{center}
\end{table*}

\section{Summary} \label{summary}

In this paper, we have established a framework to investigate the effects of host galaxy properties on SNe Ia from the DES. We have constructed seeing-optimised image stacks, free from SN light, and used them to measure both global host galaxy fluxes, and those measured locally (4\,kpc radius apertures) at the SN position. We used these data to estimate stellar masses and rest-frame $U-R$ colours using galaxy SED fitting, and compared these with the SN Ia light curve properties and luminosities. Our principal findings are:

\begin{itemize}

    \item All the measured steps are significant at $>3\sigma$ (range of 3.3--5.5$\sigma$), whether using local or global measures, or using stellar mass or $U-R$ colour, or splitting at the environmental property sample median or maximal step point.

    \item Local stellar mass steps are larger than  global stellar mass steps by up to $0.03$\,mag, and thus may recover more residual SN Ia magnitude dispersion. 

    \item Both global $U-R$ ($0.081 \pm 0.017$\,mag) and local $U-R$ ($0.082 \pm 0.017$\,mag) steps are larger than the global mass step ($0.057 \pm 0.017$\,mag). Although the difference between global and local $U-R$ steps is small, the size of the local $U-R$ step is more stable when considering different values to divide the SN sample, and thus may be less susceptible to analysis choices. 
    
    \item SNe Ia in redder (and presumably passive or dustier) galaxies have a higher r.m.s. scatter in their Hubble residuals, suggesting that SNe Ia in bluer galaxies provide a more homogeneous sample.
   
    \item When we split our SN Ia sample by the SN colour $c$, we find results that do not agree with earlier studies by \citep[e.g.,][]{2010MNRAS.406..782S,2018arXiv180603849R}. We find the redder objects ($c > 0$) have larger steps for both stellar mass and $U-R$, for both global and local, of $\sim 0.14$\,mag. 
    
    \item The homogeneity for SNe Ia in bluer galaxies and environments is most pronounced when splitting into sub-samples based on $c$, with an r.m.s. scatter of $0.084 \pm 0.017$\,mag for SNe Ia in bluer local environments when $c < 0$. 
   
    \item When we split our sample by environmental property and refit the nuisance parameters $\alpha$ and $\beta$, we find mild tension ($\sim 2\sigma$ difference) in $\alpha$ and $\beta$ across the division point. Smaller $\beta$ values are observed in higher mass, redder regions (or galaxies), agreeing with the prediction of \citet{2020arXiv200410206B}. We also find a smaller $\alpha$ for low mass, bluer regions, suggesting that the most cosmologically-uniform sample is in actively star-forming, lower-mass galaxies. 
    
\end{itemize}

These results have implications for using SNe Ia as cosmological probes. However, the DESYR3 sample that we consider here, despite its exquisite photometric calibration and the spectroscopic confirmation of all SNe in the sample, remains modest in size, particularly after segregating the sample by SN light-curve parameters. The upcoming DES-5YR sample of SNe Ia will be significantly larger, and thus provide further insight in understanding the effect of environment on SNe Ia.

\section*{Acknowledgements}

This work was supported by the Science and Technology Facilities Council [grant number ST/P006760/1] through the DISCnet Centre for Doctoral Training. M.Sullivan and M.Smith acknowledge support from EU/FP7-ERC grant 615929, and P.W. acknowledges support from STFC grant ST/R000506/1. L.G. was funded by the European Union's Horizon 2020 research and innovation programme under the Marie Sk\l{}odowska-Curie grant agreement No. 839090. This work has been partially supported by the Spanish grant PGC2018-095317-B-C21 within the European Funds for Regional Development (FEDER).

This research made use of Astropy,\footnote{\url{http://www.astropy.org}} a community-developed core Python package for Astronomy \citep{astropy:2013, astropy:2018}. This research made use of Photutils, an Astropy package for detection and photometry of astronomical sources \citep{larry_bradley_2019_2533376}.

%DES stuff
Funding for the DES Projects has been provided by the U.S. Department of Energy, the U.S. National Science Foundation, the Ministry of Science and Education of Spain, 
the Science and Technology Facilities Council of the United Kingdom, the Higher Education Funding Council for England, the National Center for Supercomputing 
Applications at the University of Illinois at Urbana-Champaign, the Kavli Institute of Cosmological Physics at the University of Chicago, 
the Center for Cosmology and Astro-Particle Physics at the Ohio State University,
the Mitchell Institute for Fundamental Physics and Astronomy at Texas A\&M University, Financiadora de Estudos e Projetos, 
Funda{\c c}{\~a}o Carlos Chagas Filho de Amparo {\`a} Pesquisa do Estado do Rio de Janeiro, Conselho Nacional de Desenvolvimento Cient{\'i}fico e Tecnol{\'o}gico and 
the Minist{\'e}rio da Ci{\^e}ncia, Tecnologia e Inova{\c c}{\~a}o, the Deutsche Forschungsgemeinschaft and the Collaborating Institutions in the Dark Energy Survey. 

The Collaborating Institutions are Argonne National Laboratory, the University of California at Santa Cruz, the University of Cambridge, Centro de Investigaciones Energ{\'e}ticas, 
Medioambientales y Tecnol{\'o}gicas-Madrid, the University of Chicago, University College London, the DES-Brazil Consortium, the University of Edinburgh, 
the Eidgen{\"o}ssische Technische Hochschule (ETH) Z{\"u}rich, 
Fermi National Accelerator Laboratory, the University of Illinois at Urbana-Champaign, the Institut de Ci{\`e}ncies de l'Espai (IEEC/CSIC), 
the Institut de F{\'i}sica d'Altes Energies, Lawrence Berkeley National Laboratory, the Ludwig-Maximilians Universit{\"a}t M{\"u}nchen and the associated Excellence Cluster Universe, 
the University of Michigan, the National Optical Astronomy Observatory, the University of Nottingham, The Ohio State University, the University of Pennsylvania, the University of Portsmouth, 
SLAC National Accelerator Laboratory, Stanford University, the University of Sussex, Texas A\&M University, and the OzDES Membership Consortium.

Based in part on observations at Cerro Tololo Inter-American Observatory, National Optical Astronomy Observatory, which is operated by the Association of 
Universities for Research in Astronomy (AURA) under a cooperative agreement with the National Science Foundation.

The DES data management system is supported by the National Science Foundation under Grant Numbers AST-1138766 and AST-1536171.
The DES participants from Spanish institutions are partially supported by MINECO under grants AYA2015-71825, ESP2015-66861, FPA2015-68048, SEV-2016-0588, SEV-2016-0597, and MDM-2015-0509, 
some of which include ERDF funds from the European Union. IFAE is partially funded by the CERCA program of the Generalitat de Catalunya.
Research leading to these results has received funding from the European Research
Council under the European Union's Seventh Framework Program (FP7/2007-2013) including ERC grant agreements 240672, 291329, and 306478.
We  acknowledge support from the Brazilian Instituto Nacional de Ci\^encia
e Tecnologia (INCT) e-Universe (CNPq grant 465376/2014-2).

This manuscript has been authored by Fermi Research Alliance, LLC under Contract No. DE-AC02-07CH11359 with the U.S. Department of Energy, Office of Science, Office of High Energy Physics.

%%%%%%%%%%%%%%%%%%%%%%%%%%%%%%%%%%%%%%%%%%%%%%%%%%
\section*{Data Availability Statement}

The data underlying this article are available in the DES3YR data release, available at \url{https://www.darkenergysurvey.org/des-year-3-supernova-cosmology-results/}, and in the online supplementary material.

\bibliographystyle{mnras}
\bibliography{biblio}

% if your bibtex file is called example.bib

%%%%%%%%%%%%%%%%%%%%%%%%%%%%%%%%%%%%%%%%%%%%%%%%%%
%%%%%%%%%%%%%%%%% APPENDICES %%%%%%%%%%%%%%%%%%%%%
%\appendix 
%%%%%%%%%%%%%%%%%%%%%%%%%%%%%%%%%%%%%%%%%%%%%%%%%%

\clearpage
\onecolumn
%\parbox{\textwidth}{
%\scriptsize
\noindent
$^{1}$ School of Physics and Astronomy, University of Southampton,  Southampton, SO17 1BJ, UK\\
$^{2}$ DISCnet Centre for Doctoral Training, University of Southampton, Southampton SO17 1BJ, United Kingdom\\
$^{3}$ Universit\'e de Lyon, F-69622, Lyon, France; Universit\'e de Lyon 1, Villeurbanne; CNRS/IN2P3, Institut de Physique des Deux Infinis, Lyon\\
$^{4}$ Department of Physics and Astronomy, University of Pennsylvania, Philadelphia, PA 19104, USA\\
$^{5}$ NASA Einstein Fellow\\
$^{6}$ Center for Astrophysics $\vert$ Harvard \& Smithsonian, 60 Garden Street, Cambridge, MA 02138, USA\\
$^{7}$ School of Mathematics and Physics, University of Queensland,  Brisbane, QLD 4072, Australia\\
$^{8}$ Institute of Cosmology and Gravitation, University of Portsmouth, Portsmouth, PO1 3FX, UK\\
$^{9}$ Departamento de F\'isica Te\'orica y del Cosmos, Universidad de Granada, E-18071 Granada, Spain\\
$^{10}$ Department of Astronomy and Astrophysics, University of Chicago, Chicago, IL 60637, USA\\
$^{11}$ Kavli Institute for Cosmological Physics, University of Chicago, Chicago, IL 60637, USA\\
$^{12}$ Centre for Gravitational Astrophysics, College of Science, The Australian National University, ACT 2601, Australia\\
$^{13}$ The Research School of Astronomy and Astrophysics, Australian National University, ACT 2601, Australia\\
$^{14}$ Universit\'e Clermont Auvergne, CNRS/IN2P3, LPC, F-63000 Clermont-Ferrand, France\\
$^{15}$ Department of Physics, Duke University Durham, NC 27708, USA\\
$^{16}$ McDonald Observatory, The University of Texas at Austin, Fort Davis, TX 79734\\
$^{17}$ Cerro Tololo Inter-American Observatory, NSF's National Optical-Infrared Astronomy Research Laboratory, Casilla 603, La Serena, Chile\\
$^{18}$ Departamento de F\'isica Matem\'atica, Instituto de F\'isica, Universidade de S\~ao Paulo, CP 66318, S\~ao Paulo, SP, 05314-970, Brazil\\
$^{19}$ Laborat\'orio Interinstitucional de e-Astronomia - LIneA, Rua Gal. Jos\'e Cristino 77, Rio de Janeiro, RJ - 20921-400, Brazil\\
$^{20}$ Fermi National Accelerator Laboratory, P. O. Box 500, Batavia, IL 60510, USA\\
$^{21}$ Instituto de Fisica Teorica UAM/CSIC, Universidad Autonoma de Madrid, 28049 Madrid, Spain\\
$^{22}$ CNRS, UMR 7095, Institut d'Astrophysique de Paris, F-75014, Paris, France\\
$^{23}$ Sorbonne Universit\'es, UPMC Univ Paris 06, UMR 7095, Institut d'Astrophysique de Paris, F-75014, Paris, France\\
$^{24}$ Department of Physics \& Astronomy, University College London, Gower Street, London, WC1E 6BT, UK\\
$^{25}$ Kavli Institute for Particle Astrophysics \& Cosmology, P. O. Box 2450, Stanford University, Stanford, CA 94305, USA\\
$^{26}$ SLAC National Accelerator Laboratory, Menlo Park, CA 94025, USA\\
$^{27}$ Instituto de Astrofisica de Canarias, E-38205 La Laguna, Tenerife, Spain\\
$^{28}$ Universidad de La Laguna, Dpto. Astrofísica, E-38206 La Laguna, Tenerife, Spain\\
$^{29}$ Department of Astronomy, University of Illinois at Urbana-Champaign, 1002 W. Green Street, Urbana, IL 61801, USA\\
$^{30}$ National Center for Supercomputing Applications, 1205 West Clark St., Urbana, IL 61801, USA\\
$^{31}$ Institut de F\'{\i}sica d'Altes Energies (IFAE), The Barcelona Institute of Science and Technology, Campus UAB, 08193 Bellaterra (Barcelona) Spain\\
$^{32}$ Institut d'Estudis Espacials de Catalunya (IEEC), 08034 Barcelona, Spain\\
$^{33}$ Institute of Space Sciences (ICE, CSIC),  Campus UAB, Carrer de Can Magrans, s/n,  08193 Barcelona, Spain\\
$^{34}$ INAF-Osservatorio Astronomico di Trieste, via G. B. Tiepolo 11, I-34143 Trieste, Italy\\
$^{35}$ Institute for Fundamental Physics of the Universe, Via Beirut 2, 34014 Trieste, Italy\\
$^{36}$ Observat\'orio Nacional, Rua Gal. Jos\'e Cristino 77, Rio de Janeiro, RJ - 20921-400, Brazil\\
$^{37}$ Department of Physics, IIT Hyderabad, Kandi, Telangana 502285, India\\
$^{38}$ Santa Cruz Institute for Particle Physics, Santa Cruz, CA 95064, USA\\
$^{39}$ Institute of Theoretical Astrophysics, University of Oslo. P.O. Box 1029 Blindern, NO-0315 Oslo, Norway\\
$^{40}$ Jet Propulsion Laboratory, California Institute of Technology, 4800 Oak Grove Dr., Pasadena, CA 91109, USA\\
$^{41}$ Department of Astronomy, University of Michigan, Ann Arbor, MI 48109, USA\\
$^{42}$ Department of Physics, University of Michigan, Ann Arbor, MI 48109, USA\\
$^{43}$ Department of Physics, Stanford University, 382 Via Pueblo Mall, Stanford, CA 94305, USA\\
$^{44}$ Center for Cosmology and Astro-Particle Physics, The Ohio State University, Columbus, OH 43210, USA\\
$^{45}$ Department of Physics, The Ohio State University, Columbus, OH 43210, USA\\
$^{46}$ Lawrence Berkeley National Laboratory, 1 Cyclotron Road, Berkeley, CA 94720, USA\\
$^{47}$ Australian Astronomical Optics, Macquarie University, North Ryde, NSW 2113, Australia\\
$^{48}$ Lowell Observatory, 1400 Mars Hill Rd, Flagstaff, AZ 86001, USA\\
$^{49}$ George P. and Cynthia Woods Mitchell Institute for Fundamental Physics and Astronomy, and Department of Physics and Astronomy, Texas A\&M University, College Station, TX 77843,  USA\\
$^{50}$ Department of Astronomy, The Ohio State University, Columbus, OH 43210, USA\\
$^{51}$ Radcliffe Institute for Advanced Study, Harvard University, Cambridge, MA 02138\\
$^{52}$ Instituci\'o Catalana de Recerca i Estudis Avan\c{c}ats, E-08010 Barcelona, Spain\\
$^{53}$ Physics Department, 2320 Chamberlin Hall, University of Wisconsin-Madison, 1150 University Avenue Madison, WI  53706-1390\\
$^{54}$ Institute of Astronomy, University of Cambridge, Madingley Road, Cambridge CB3 0HA, UK\\
$^{55}$ Department of Astrophysical Sciences, Princeton University, Peyton Hall, Princeton, NJ 08544, USA\\
$^{56}$ Department of Physics and Astronomy, Pevensey Building, University of Sussex, Brighton, BN1 9QH, UK\\
$^{57}$ Centro de Investigaciones Energ\'eticas, Medioambientales y Tecnol\'ogicas (CIEMAT), Madrid, Spain\\
$^{58}$ Computer Science and Mathematics Division, Oak Ridge National Laboratory, Oak Ridge, TN 37831\\
$^{59}$ Max Planck Institute for Extraterrestrial Physics, Giessenbachstrasse, 85748 Garching, Germany\\
$^{60}$ Universit\"ats-Sternwarte, Fakult\"at f\"ur Physik, Ludwig-Maximilians Universit\"at M\"unchen, Scheinerstr. 1, 81679 M\"unchen, Germany\\
%}

% Don't change these lines
\bsp	% typesetting comment
\label{lastpage}
\end{document}